%% file: main.tex
\documentclass[12pt]{article}
\usepackage[a4paper,left=15mm,right=15mm,top=18mm,bottom=18mm]{geometry}
\usepackage{amsmath,graphicx,caption,booktabs,array,titlesec,fancyhdr,setspace}
\usepackage[nospace]{cite}
\usepackage[hidelinks]{hyperref}
\usepackage[T1]{fontenc}
\usepackage{placeins}
\titleformat{\section}{\fontsize{14}{17}\selectfont\bfseries}{}{0pt}{}
\titleformat{\subsection}{\fontsize{12}{15}\selectfont\bfseries}{}{0pt}{}
\titlespacing*{\section}{0pt}{11pt}{5pt}
\titlespacing*{\subsection}{0pt}{9pt}{4pt}
\makeatletter
\renewcommand{\@cite}[2]{\textsuperscript{#1}}
\renewcommand{\@biblabel}[1]{#1.}
\makeatother
\hypersetup{hypertexnames=false}
\AtBeginDocument{}
\begin{document}

\input{main-body.tex}

\clearpage
\setcounter{page}{1}
\renewcommand{\thepage}{S\arabic{page}}
\normalfont\fontencoding{T1}\selectfont

\input{supplement-body.tex}

\end{document}

%% file: main-body.tex
\hypersetup{pdfauthor={Ruiqi Dong; Haotian Huang; Yubin Hu; Guixiang Zeng}}

{\raggedright\noindent\fontsize{16}{19}\selectfont\bfseries Accelerating interatomic potential development through student-driven exploration and reliability-aware distillation\par}\vspace{4mm}

\noindent Ruiqi Dong\textsuperscript{1}, Haotian Huang\textsuperscript{1,2}, Yubin Hu\textsuperscript{1}, Guixiang Zeng\textsuperscript{1*}\par

\noindent\textsuperscript{1}Kuang Yaming Honors School, Nanjing University, Nanjing 210023, China\par

\noindent\textsuperscript{2}School of Chemistry, Nanjing University, Nanjing 210023, China\par

\noindent\textsuperscript{*}Correspondence: gxzeng@nju.edu.cn\par\vspace{4mm}

\section*{Abstract}

\noindent Conventional two-stage distillation first establishes a teacher through teacher-driven configuration exploration, a costly first stage for complex materials, and then uses its predictions to supervise students, potentially propagating teacher-label errors. Here, we propose distillation-driven teacher--student co-evolution (DICE), which uses a fast student to accelerate configuration exploration and replaces the fixed teacher-to-student sequence with alternating student and teacher updates. A representation-based support-assessment algorithm routes high-support configurations to teacher labelling and selects low-support configurations for density functional theory (DFT) calculations to update the teacher, thereby reducing the risk of teacher-label error propagation to students. Across Al\textsubscript{2}O\textsubscript{3} (structural ceramics), yttria-stabilised zirconia (solid electrolytes) and MIL-120(Al) (a metal--organic framework), DICE develops reliable compact potentials using only 140--494 target DFT reference configurations per system; total recorded student exploration times per system are just 2.6--5.1\% of the corresponding teacher-driven estimates. Combining computational efficiency with broad applicability across diverse material systems, DICE provides a practical route to developing reliable compact interatomic potentials.\par

\section*{Introduction}

\noindent Molecular dynamics (MD) simulations reveal atomic-scale structural evolution and material behaviour, and their reliability depends on accurate interatomic interactions. Density functional theory (DFT)-based ab initio molecular dynamics (AIMD) obtains atomic forces from electronic-structure calculations, but its high computational cost limits accessible spatial and temporal scales. Machine-learning interatomic potentials (MLIPs) learn from first-principles energies and forces to extend these scales with near-first-principles accuracy at much lower cost \cite{dice_ref_01,behler2007,zhang2018dpmd}. Developing reliable MLIPs nevertheless requires training data covering the chemical environments relevant to the intended simulations, and DFT labelling remains expensive.\par

Active learning reduces the DFT data requirements for training interatomic potentials through configuration exploration and selective labelling of informative configurations \cite{active2017,concurrent2019,flare2020}. Nevertheless, training from scratch still requires a substantial number of DFT references to achieve sufficient accuracy. Fine-tuning large pretrained atomistic models offers a further reduction in target-system DFT requirements by transferring learned knowledge of atomic interactions \cite{dice_ref_02,dice_ref_03,dice_ref_06}. However, their high inference costs constrain extensive configuration exploration and large-scale, long-timescale simulations. Knowledge distillation expands the student training set with teacher-generated labels to train faster compact students \cite{dice_ref_06,dice_ref_04,dice_ref_05,hinton2015,kelvinius2023distill,amin2025hessian,lim2026ark,gardner2025distillation,li2025lightpfp}. Compared with training identical student architectures directly on the same DFT references, distillation can improve predictive accuracy, dynamical stability and physical-property prediction \cite{dice_ref_05,gardner2025distillation,chen2026condensed_distillation}.\par

Conventional two-stage distillation first adapts a teacher through configuration exploration and only then uses it to supervise a student, leaving the initial exploration to the computationally expensive teacher. To reduce this cost, Zhang \textit{et al.} introduced a one-shot high-temperature MACE-MD exploration strategy for solid-state electrolytes, followed by distillation from MACE to NEP \cite{dice_ref_05}. However, single-pass exploration may fail to capture configurations relevant to specific target properties, as reported for sulfur-vacancy migration in MoS\textsubscript{2} and solvated phenol \cite{lian2026active_distillation}, as well as a choline chloride--citric acid solution containing Co\textsuperscript{2+} and Li\textsuperscript{+} \cite{haenseroth2026iterative}. Lian and Pasquarello subsequently incorporated student-driven active learning into the second stage of MACE-to-DeepMD distillation, using the adapted MACE model to label configurations explored by the DeepMD student \cite{wong2026bias}. Taken together, these efforts leave a structural limitation of two-stage distillation unresolved: an initial teacher must still be established before the student can contribute to exploration, leaving the first stage dependent on the more computationally expensive teacher model.\par

Beyond exploration costs, distillation requires assessing teacher-label accuracy because teacher errors can affect student accuracy \cite{gardner2025distillation,dice_ref_11}. For example, Gardner \textit{et al.} observed unphysical reaction products in distilled-student simulations of an $S_{\mathrm{N}}2$ reaction, which they attributed to the teacher's inaccurate description of high-energy configurations \cite{gardner2025distillation}. Uncertainty-aware model distillation (UAMD) uses a heterogeneous ensemble of unchanged pretrained teachers to assess uncertainty, applying DFT labels to selected high-uncertainty configurations and teacher labels to the remainder \cite{dice_ref_11}. This reliability-based routing reduces DFT labelling requirements. Building on this approach, we further investigate whether adding the same selected DFT references to the teacher training set and updating the teacher can improve student accuracy more effectively than directly replacing the corresponding teacher-generated labels in the student training set.\par

Here, we propose distillation-driven teacher--student co-evolution (DICE), an active-learning distillation workflow coupling fast student-driven configuration exploration with alternating student and teacher updates. Our representation-based support-assessment algorithm routes high-support configurations to teacher labelling to expand the student training set, while selected low-support configurations undergo DFT calculations to expand the teacher training set and update the teacher. OMat24-pretrained DPA4-Neo and DPA4C-Nano provide the starting points for teacher and student fine-tuning, respectively \cite{dice_ref_14,dice_ref_15,dice_ref_16}.\par

We develop compact potentials for Al\textsubscript{2}O\textsubscript{3}, yttria-stabilised zirconia (YSZ) and MIL-120(Al) using 140, 195 and 494 target DFT reference configurations, respectively (Supplementary Tables 1--3). For equivalent exploration workloads, total recorded student-driven exploration times per system are only approximately 2.6--5.1\% of the corresponding teacher-driven estimates (Supplementary Table 4). The resulting potentials predict structural, energetic and dynamical properties consistently with DFT calculations and experiments. DICE thus brings compact-model acceleration into continuing potential development and, through teacher adaptation, extends the benefit of selected DFT references to additional student-training configurations.\par

\section*{Results}

\FloatBarrier

\subsection*{Overview of the DICE workflow}

\noindent DICE uses a fast student to drive configuration exploration and alternates student and teacher updates within a single workflow (Fig. 1). Within each iteration, the teacher remains unchanged while student training, configuration exploration and support-guided routing repeat in cycles indexed r0, r1, r2, \ldots{}. DIRECT-selected high-support configurations receive teacher labels to expand the student training set, and the updated student undertakes the next exploration cycle; low-support configurations accumulate in a DFT-labelling candidate pool. At teacher updates, DFT calculations on selected candidates expand the teacher training set. The updated teacher and rebuilt support estimator are then used in the next iteration. This routing uses our representation-based support-assessment algorithm, which fits Gaussian mixture models (GMMs) to atomic representations from the teacher training set and requires a single teacher. Two design choices underpin this workflow: using selective DFT feedback to update the teacher and using single-teacher support assessment to select configurations for teacher labelling.\par

\begin{figure}[!htbp]\centering
\includegraphics[width=180.00mm,height=118.85mm,keepaspectratio]{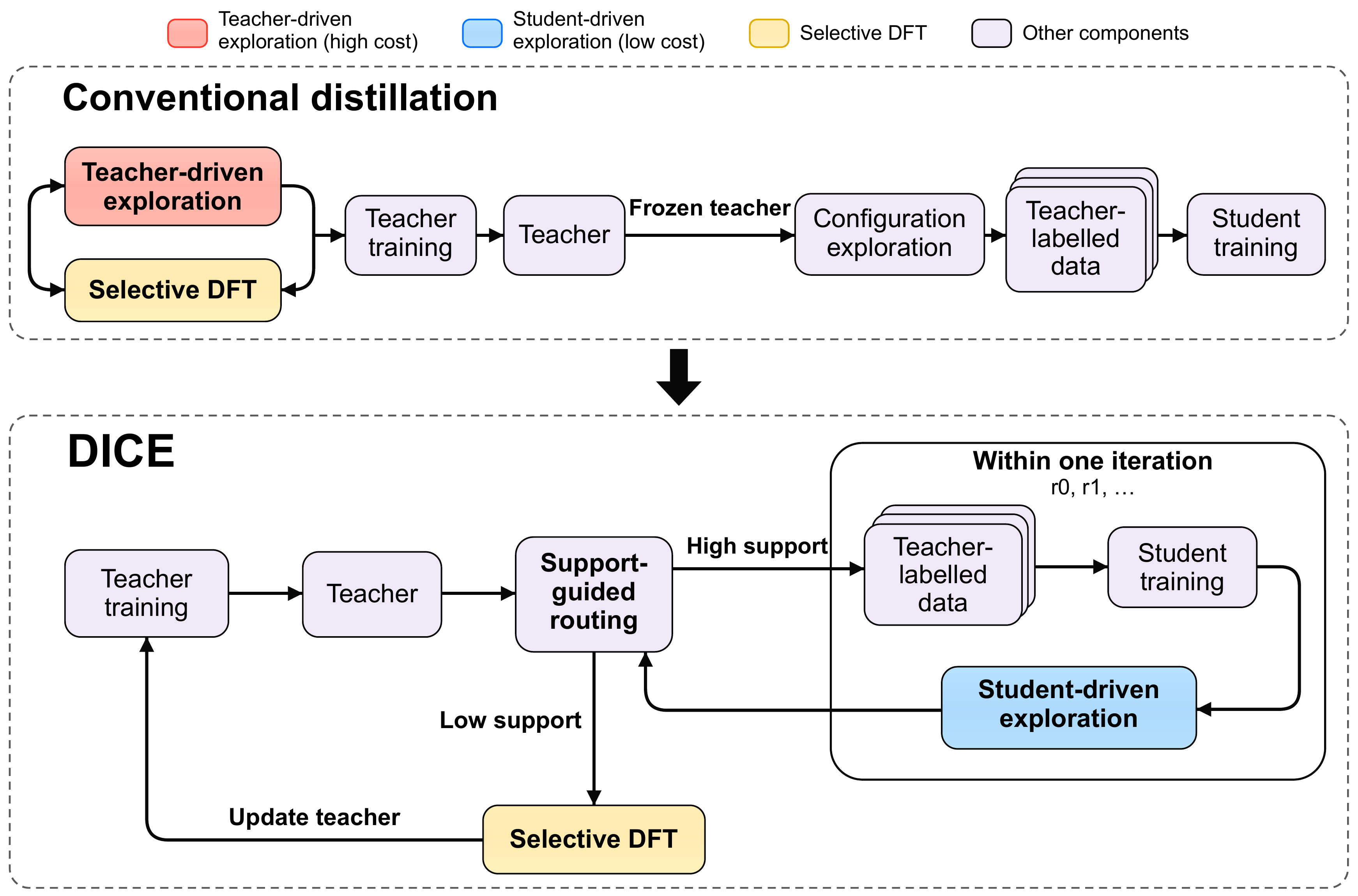}
\par\vspace{2mm}{\fontsize{10}{12}\selectfont\raggedright\textbf{Fig. 1. Conventional distillation and the DICE workflow.} Top, teacher-driven exploration and selective DFT establish a teacher, which is then frozen for student supervision. Bottom, DICE uses student-driven exploration: DIRECT-selected high-support configurations receive teacher labels to expand the student training set; selective DFT feedback from low-support configurations expands the teacher training set and updates the teacher.\par}
\end{figure}

\FloatBarrier

\subsection*{Updating the teacher with DFT data outperforms replacing selected labels}

\noindent We first use controlled experiments to examine whether teacher-generated labels yield students as accurate as those trained on DFT labels and, when a gap arises, how additional DFT data can more effectively improve student accuracy. The experiments use DFT-labelled crystalline $\gamma$-Smr\v{c}ok and liquid--amorphous Al\textsubscript{2}O\textsubscript{3} datasets containing 516 and 423 configurations \cite{dice_ref_33,dice_ref_34}. We fine-tune pretrained DPA4-Neo on ten $\gamma$-Smr\v{c}ok references to obtain the initial teacher $T_0$, then randomly select sixty configurations from the remaining $\gamma$-Smr\v{c}ok pool or the liquid--amorphous dataset and combine them with the initial ten. Students are fine-tuned on these seventy training configurations with $T_0$ or DFT labels, the latter denoted DFT-70. All student routes start from the same pretrained DPA4C-Nano model, use identical fine-tuning settings and are evaluated on common held-out configurations within each repeat.\par

On $\gamma$-Smr\v{c}ok, students trained on $T_0$ labels closely match DFT-70, whereas on liquid--amorphous configurations their errors are markedly higher (Fig. 2a; Supplementary Fig. 1c,d). This difference corresponds to $T_0$'s lower force errors on $\gamma$-Smr\v{c}ok and higher errors on liquid--amorphous configurations (Fig. 2b), showing how teacher-label errors affect student accuracy relative to fully DFT-labelled training.\par

\begin{figure}[!htbp]\centering
\includegraphics[width=180.00mm,height=99.35mm,keepaspectratio]{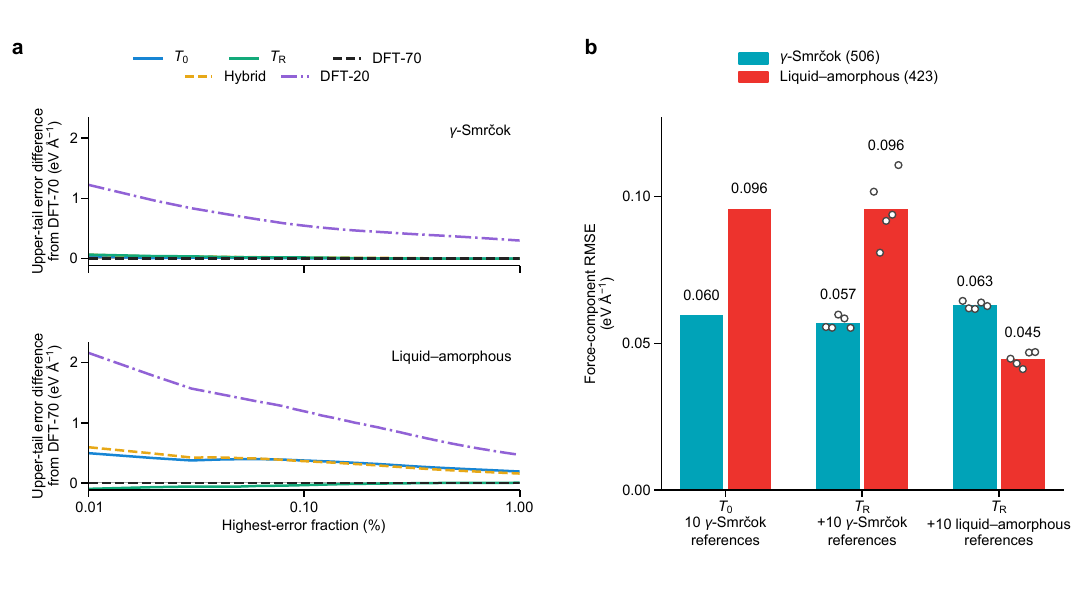}
\par\vspace{2mm}{\fontsize{10}{12}\selectfont\raggedright\textbf{Fig. 2. Updating the teacher with additional DFT references more effectively improves student supervision.} a, Differences in students' upper-tail force errors (route minus DFT-70) on held-out $\gamma$-Smr\v{c}ok (upper) and liquid--amorphous (lower) configurations. Curves show mean paired differences across five configuration-selection repeats; tails average errors within the indicated highest-error fraction. Routes use initial ($T_0$) or updated ($T_R$) teacher labels, teacher labels replaced by DFT labels for ten queried configurations (Hybrid), or twenty or seventy DFT references (DFT-20 or DFT-70). b, Teacher force-component RMSE on the remaining $\gamma$-Smr\v{c}ok pool (506; excluding the initial ten references) and the full liquid--amorphous pool (423), including newly queried configurations. Bars show means; points show five updated teachers; $T_0$ is evaluated once.\par}
\end{figure}

To narrow the liquid--amorphous student accuracy gap from DFT-70, we randomly select ten of the sixty additions in each group as new DFT references. Drawing on UAMD \cite{dice_ref_11}, Hybrid replaces labels on these ten configurations with DFT labels and retains $T_0$ labels on the other sixty, including the initial ten. With the seventy training configurations unchanged, Hybrid lowers overall student error but retains a clear gap from DFT-70 (Fig. 2a; Supplementary Fig. 1c,d).\par

Using the same new DFT references to update the teacher improves student accuracy more effectively. We combine them with the initial ten, fine-tune $T_R$ from the same pretrained DPA4-Neo model and relabel all seventy training configurations. On held-out liquid--amorphous configurations, updating the teacher with the same additional DFT references narrows the gap in mean student force-component RMSE relative to DFT-70 from approximately 7.5\% for Hybrid to 0.2\% (Supplementary Fig. 1c,d). The upper-tail error comparison likewise shows the advantage of teacher updating (Fig. 2a). In contrast, DFT-20, fine-tuned only on the same twenty DFT references, has higher errors, showing that extra teacher-labelled configurations further improve student accuracy under the same DFT budget (Fig. 2a).\par

Teacher updates improve predictions on queried and unqueried configurations. In the liquid--amorphous experiments, updating the teacher reduces overall error across the seventy student-training configurations and improves predictions on the fifty unqueried additions and the held-out configurations (Supplementary Fig. 1a,b). Additional DFT references thus provide more accurate labels for further student-training configurations through teacher updates.\par

Reliable distillation should yield students as accurate as those trained entirely on DFT labels for the same configurations under identical fine-tuning settings. When teacher-label errors create a student accuracy gap relative to this benchmark, using additional DFT references to update the teacher is more effective than replacing the corresponding labels and extends their benefit beyond directly queried configurations.\par

\FloatBarrier

\subsection*{Support-guided routing enables reliable teacher labelling}

\noindent The preceding experiments show that differences in teacher-label quality across configurations affect student accuracy. Expanding the student training set with teacher labels therefore requires identifying configurations that can receive reliable teacher labels. We assess whether routed supervision maintains student accuracy while expanding the training set, using students trained entirely on DFT labels for the same geometries as controls.\par

We first examine query-by-committee (QBC) \cite{smith2018active} in the $\gamma$-Smr\v{c}ok and liquid--amorphous Al\textsubscript{2}O\textsubscript{3} pools. Four teachers are fine-tuned from the same pretrained DPA4-Neo model on the initial ten $\gamma$-Smr\v{c}ok references with different random seeds. The relationship between maximum atomic force disagreement $\sigma$ and the prediction error of the labelling teacher $T_0$ differs markedly between the pools: liquid--amorphous configurations exhibit large errors even at low $\sigma$ (Fig. 3a,b). QBC accepts configurations for teacher labelling when $\sigma$ does not exceed the threshold. Following the lower force-disagreement threshold used for active-learning selection in previous Al\textsubscript{2}O\textsubscript{3} work \cite{qbc_yu2026alumina}, we test 0.08 eV Å\textsuperscript{\ensuremath{-}1} and the stricter thresholds of 0.05 and 0.03 eV Å\textsuperscript{\ensuremath{-}1}. Defining high error as per-configuration force-component RMSE > 0.05 eV Å\textsuperscript{\ensuremath{-}1}, even the strictest threshold accepts a set containing 27.8\% high-error liquid--amorphous configurations, compared with 0.67\% in $\gamma$-Smr\v{c}ok (Supplementary Fig. 2a,b). The same $\sigma$ threshold therefore yields markedly different teacher-label quality.\par

\begin{figure}[!htbp]\centering
\includegraphics[width=180.00mm,height=151.48mm,keepaspectratio]{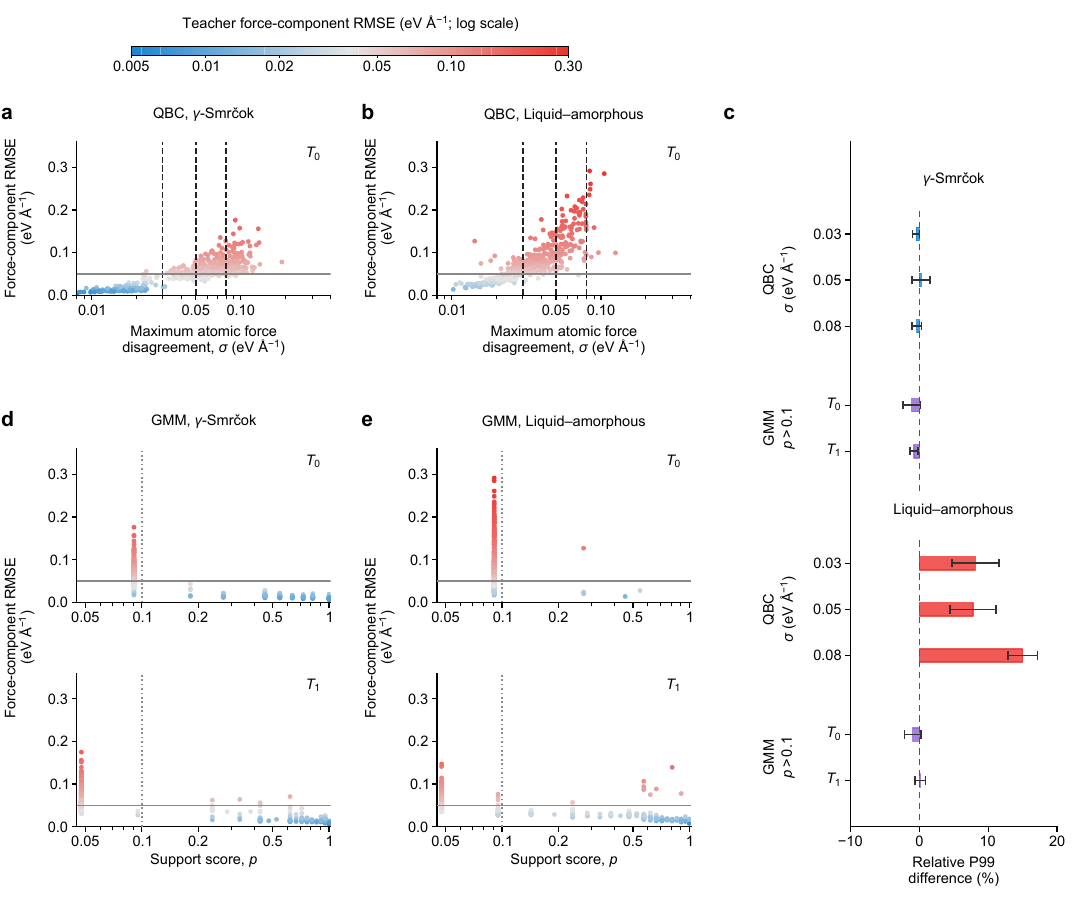}
\par\vspace{2mm}{\fontsize{10}{12}\selectfont\raggedright\textbf{Fig. 3. Support-guided routing selects configurations for reliable teacher labelling.} a,b, QBC force disagreement versus initial-teacher force-component RMSE in $\gamma$-Smr\v{c}ok (a) and liquid--amorphous (b) pools. c, Percentage differences in the P99 of per-atom force-error magnitudes for students trained on teacher labels relative to students trained entirely on DFT labels, using identical training geometries (mean \ensuremath{\pm} sample SD, five repeats). d,e, Support score $p$ versus teacher RMSE in the same pools before (top, $T_0$) and after (bottom, $T_1$) updating; updated-teacher scatters show the first configuration-selection repeat. Colours in a,b,d,e show RMSE; vertical and horizontal lines mark routing thresholds and the 0.05 eV Å\textsuperscript{\ensuremath{-}1} error criterion.\par}
\end{figure}

This difference in label quality is also reflected in student accuracy. We randomly select fifty accepted configurations, using all when fewer are available, combine them with the initial ten references and compare $T_0$ and DFT labels on the same training geometries. Student accuracy is similar between label sources in $\gamma$-Smr\v{c}ok, whereas students trained on teacher labels have higher errors in the liquid--amorphous comparisons at all three thresholds (Supplementary Fig. 2c,d), with the 99th percentile (P99) of per-atom force-error magnitudes averaging 7.8--15.0\% above the matched controls trained entirely on DFT labels (Fig. 3c). A QBC threshold suitable for one distribution can therefore accept label errors that impair student accuracy in another. Previous studies use DFT reference labels to calibrate the relationship between committee disagreement and prediction error \cite{cezar2025cago,bidoggia2026trainspot}. Applying such methods to determine an acceptance threshold for a target distribution requires reference-label preparation and calibration analysis, introducing hidden development costs.\par

Motivated by systematic comparisons showing good GMM performance in out-of-distribution detection \cite{dice_ref_23}, we develop a GMM-based support-assessment algorithm. GMMs describe the atomic-representation distribution of the teacher training data \cite{dice_ref_26}. Candidate nonconformity scores quantify unfamiliarity in the lowest-support atomic environments; ranking them against scores of teacher training configurations defines the configuration support score $p$. A given $p$ threshold has the same tail interpretation across reference distributions, with the corresponding score boundary determined by the teacher training distribution before candidate DFT errors are available. We use 0.1 as the support threshold, accepting configurations with $p>0.1$ for teacher labelling and sending those with $p\leq0.1$ to the DFT-labelling candidate pool. With the initial ten references, the minimum $p$ is 1/11, so $p\leq0.1$ identifies candidates whose nonconformity scores exceed those of all teacher training configurations (Fig. 3d,e).\par

We test this rule in both pools using the initial teacher $T_0$ and the updated teacher $T_1$, with 0.1 as the support threshold throughout. In each pool, ten low-support DFT references expand the teacher training set to fine-tune $T_1$ from the same pretrained DPA4-Neo model. After support reassessment, we randomly add up to fifty accepted configurations; $T_1$ relabels the expanded student training set, including the new references. With $T_0$, no accepted $\gamma$-Smr\v{c}ok configuration is high-error; only five liquid--amorphous configurations are accepted, one exceeding the error threshold. The corresponding fractions in both pools are below those at all tested QBC thresholds. After teacher updating, accepted-set high-error fractions average 5.5\% and 5.9\%, respectively (Supplementary Fig. 2a,b). Matched-geometry student comparisons further assess how the selected teacher labels affect student accuracy. Across all four comparisons, covering both configuration distributions before and after teacher updating, students trained on teacher labels achieve accuracy close to that of students trained entirely on DFT labels, with mean relative P99 differences of \ensuremath{-}1.1\% to +0.1\% (Fig. 3c). The same support threshold thus selects effective supervision across distributions and teacher updates.\par

Meanwhile, the algorithm requires a single teacher, avoiding the additional cost of training and evaluating a committee. Rebuilding the support estimator after teacher updating allows continued selection of configurations for teacher labelling using a threshold with the same statistical interpretation. Based on this selection performance, explicit threshold basis and single-teacher cost advantage, we adopt the GMM-based support-assessment algorithm for configuration routing in DICE, selecting configurations for teacher labelling and directing low-support configurations to the DFT-labelling candidate pool.\par

\FloatBarrier

\subsection*{Teacher labels improve student accuracy without additional DFT}

\noindent Applying support-guided routing during DICE development for Al\textsubscript{2}O\textsubscript{3} reveals a striking contrast. The teacher is fine-tuned on 140 crystal-origin target DFT references, yet approximately 99\% of prescreened candidates submitted to support assessment during subsequent student-driven crystal-heating and low-density melt--cool exploration satisfy the high-support criterion $p>0.1$.\par

These 140 references accumulate during DICE development. Starting from twenty initial references, NVT exploration at 100--2000 K uses crystals and configurations generated by lattice strain or random atomic displacements as starting structures. Each selective DFT feedback update adds twenty references to expand the teacher training set and update the teacher, reaching 140 references by iteration 6. Within this iteration, the teacher remains unchanged while 380 DIRECT-selected high-support liquid--amorphous configurations receive teacher labels and progressively enter the student training set.\par

Structural analysis shows that these accepted additions extend the structural coverage of the student training set. Mean oxygen--oxygen partial structure factors for the teacher and pre-expansion student training sets exhibit pronounced sharp peaks, whereas the 380 additions show broader peaks resembling those of an independent 120-configuration liquid--amorphous test set (Fig. 4a). Oxygen-network bond-orientational order and normalised reciprocal-space peak intensity further show that the two existing training sets mainly occupy more ordered regions, whereas the additions extend into the less ordered region populated by the test set and include intermediate configurations connecting the more and less ordered regions (Fig. 4b).\par

\begin{figure}[!htbp]\centering
\includegraphics[width=179.55mm,height=210.00mm,keepaspectratio]{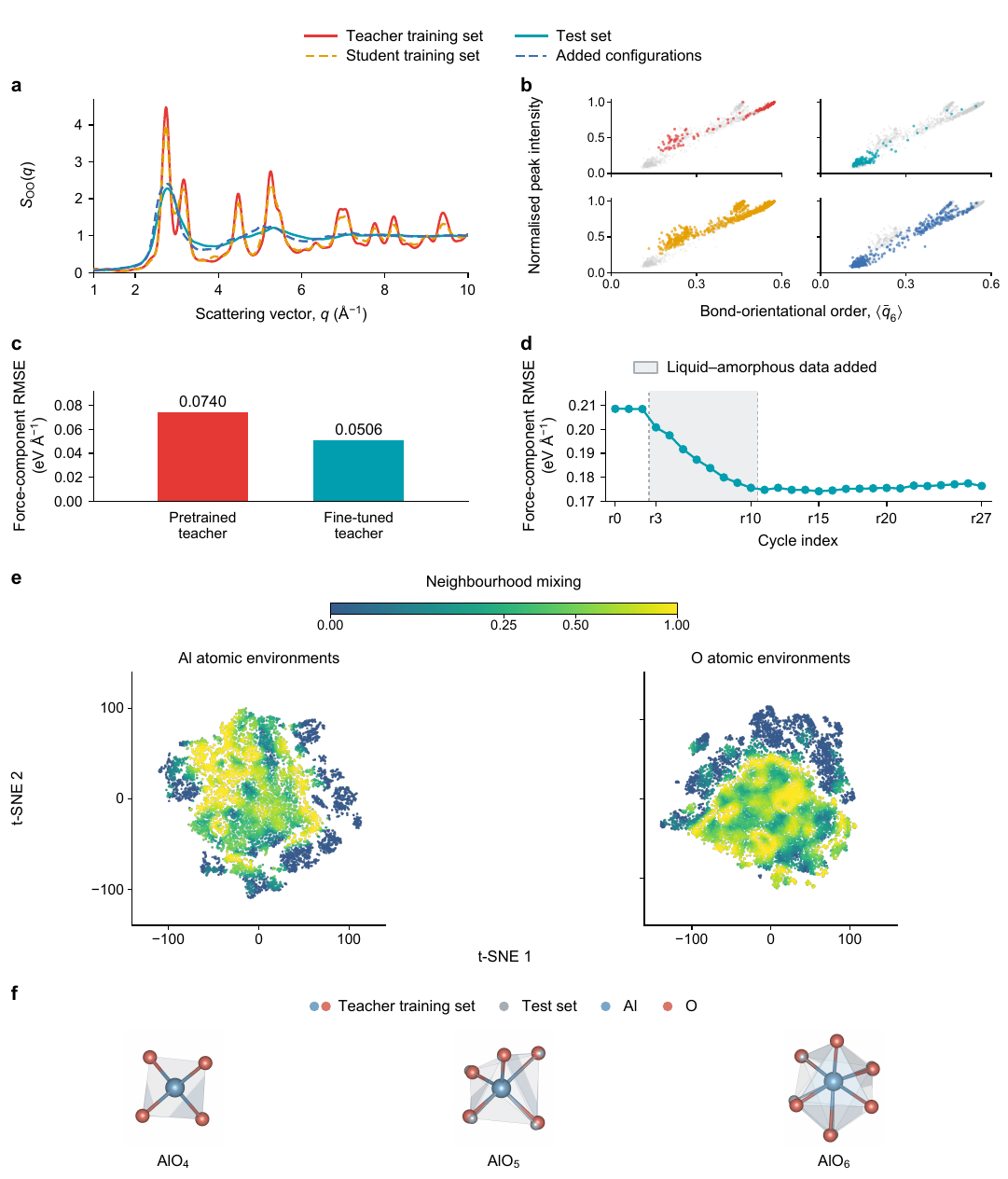}
\par\vspace{2mm}{\fontsize{10}{12}\selectfont\raggedright\textbf{Fig. 4. Support-guided routing improves student accuracy without additional DFT references.} a,b, Oxygen--oxygen partial structure factors and structural-order metrics for the teacher training set (140), pre-expansion student training set (940; r2), added configurations (380) and test set (120); grey points show all datasets. c, Overall force-component RMSEs of pretrained and fine-tuned teachers on the same test set. d, Student force-component RMSE across r0--r27 of iteration 6; shading marks additions during r3--r10. e, Joint t-SNE projections of teacher representations for Al-centred (left) and O-centred (right) environments from the teacher training and test sets. Each point represents one atomic environment; colours show source-normalised neighbourhood mixing (0, single-source; 1, balanced; square-root colour normalisation). f, Rigidly aligned AlO\textsubscript{4}, AlO\textsubscript{5} and AlO\textsubscript{6} environments from teacher training (coloured) and test (grey) sets; translucent faces show coordination polyhedra.\par}
\end{figure}

Independent testing further shows that fine-tuning on crystal-origin target DFT references improves teacher predictions for liquid--amorphous configurations. On the independent 120-configuration test set, the fine-tuned teacher's overall force-component RMSE is approximately 32\% lower than that of the pretrained model (Fig. 4c). On the high-support subset of this same test set satisfying $p>0.1$, the fine-tuned teacher's force-component RMSE is 0.0467 eV Å\textsuperscript{\ensuremath{-}1}, showing that support-guided routing identifies teacher supervision with low prediction errors.\par

These teacher labels further improve student accuracy without additional target DFT references. With student training settings unchanged, the 380 configurations selected during r2--r9 progressively enter training in r3--r10 within iteration 6. Force-component RMSE on the same independent test set remains nearly constant over r0--r2, then decreases as teacher-labelled configurations are added, reaching a reduction of approximately 16\% at r10 relative to r2. After r10, further training additions are predominantly crystalline, and liquid--amorphous test error remains nearly constant through r27 (Fig. 4d).\par

Local-environment analysis provides a structural interpretation of this teacher generalisation and the high-support assignments. For Al and O atomic environments separately, representations extracted by the fine-tuned teacher are used to construct joint t-SNE projections of the teacher training and liquid--amorphous test sets. Within each element-specific projection, the distributions of the teacher training and test sets differ overall but partially overlap (Fig. 4e). AlO\textsubscript{4}, AlO\textsubscript{5} and AlO\textsubscript{6} examples selected using a common geometric matching rule and rigidly aligned further reveal geometrically similar local coordination environments across the datasets (Fig. 4f). Both teacher predictions and support assessment use local atomic representations; these connections across phases are consistent with the observed teacher generalisation and high support. Through support-guided routing, DICE translates this generalisation into effective student-training data, allowing existing DFT references to support learning across a broader range of configurations.\par

\FloatBarrier

\subsection*{DICE distillation outperforms direct fine-tuning}

\noindent We apply DICE to potential development for Al\textsubscript{2}O\textsubscript{3}, YSZ and MIL-120(Al). Iterative learning in Al\textsubscript{2}O\textsubscript{3} shows the accuracy gains from teacher-labelled configurations during development. To evaluate how this expanded supervision improves the final potentials, we compare the final students obtained through DICE development in all three systems with students directly fine-tuned on the same target DFT references. Both routes start from the same pretrained DPA4C-Nano model and use identical fine-tuning settings. Al\textsubscript{2}O\textsubscript{3}, YSZ and MIL-120(Al) use 140, 195 and 494 target DFT reference configurations, respectively; teacher labelling expands the DICE student training sets to 2580, 4272 and 6041 configurations, approximately 12--22 times the corresponding DFT reference counts.\par

Compared with directly fine-tuned students, DICE-distilled students reduce independent-test force-component RMSE by 47.6\%, 21.9\% and 18.8\% for Al\textsubscript{2}O\textsubscript{3}, YSZ and MIL-120(Al), respectively, with both the central and higher-error regions of the per-configuration error distributions shifting towards lower errors (Fig. 5a; Supplementary Figs. 3--5). In YSZ, the fractions of O, Zr and Y atomic-force vector errors above 0.30 eV Å\textsuperscript{\ensuremath{-}1} decrease by approximately 52--88\% (Fig. 5b).\par

\begin{figure}[!htbp]\centering
\includegraphics[width=180.00mm,height=160.57mm,keepaspectratio]{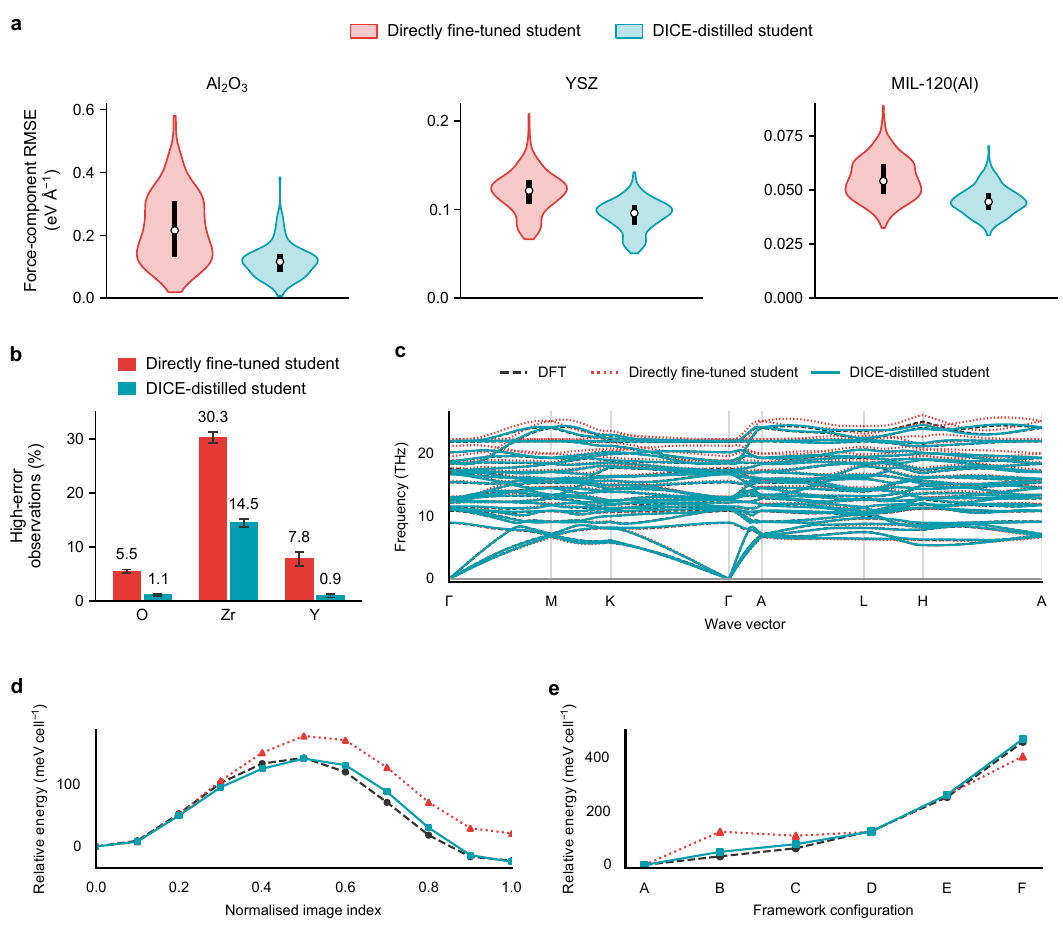}
\par\vspace{2mm}{\fontsize{10}{12}\selectfont\raggedright\textbf{Fig. 5. Practical benefits of expanded supervision over direct fine-tuning.} a, Per-configuration force-component RMSE distributions for directly fine-tuned and DICE-distilled students in Al\textsubscript{2}O\textsubscript{3}, YSZ and MIL-120(Al); white circles and lines show medians and interquartile ranges. b, YSZ element-resolved fractions of atom--configuration observations with per-atom force-error magnitudes > 0.30 eV Å\textsuperscript{\ensuremath{-}1}; error bars show 95\% paired structural-cluster bootstrap confidence intervals. c, $\alpha$-Al\textsubscript{2}O\textsubscript{3} phonon dispersions on matched displacement geometries. d, MIL-120(Al) C\ensuremath{\rightarrow}B hydroxyl-rearrangement energy profiles on common DFT geometries, with each method's initial-image energy as zero. e, EOS-fitted minimum energies of six framework configurations after internal-coordinate relaxation, with each method's configuration-A minimum as zero.\par}
\end{figure}

Improved accuracy on independent configurations is accompanied by greater dynamical stability. In MD tests of Al\textsubscript{2}O\textsubscript{3} at 2400 K, 10YSZ (nominally 10 mol\% Y\textsubscript{2}O\textsubscript{3}) at 1800 K and the MIL-120(Al) framework at 500 K, all directly fine-tuned students exhibit dynamical instability, whereas all DICE-distilled students stably complete 1 ns simulations (Supplementary Table 5).\par

Beyond prediction accuracy and dynamical stability, we further compare the accuracy of physical-property predictions from DICE-distilled and directly fine-tuned students. On matched $\alpha$-Al\textsubscript{2}O\textsubscript{3} displacement configurations, the DICE-distilled student reduces the mean absolute phonon-frequency error against DFT by approximately 79\% relative to the directly fine-tuned student (Fig. 5c). For MIL-120(Al), on the same DFT pathway configurations, the DICE-distilled student more accurately reproduces the energy profiles of the nine hydroxyl-rearrangement pathways and the relative endpoint energies of the representative C\ensuremath{\rightarrow}B pathway (Fig. 5d; Supplementary Fig. 6). Minima fitted from energy--volume curves for six MIL-120(Al) framework configurations further show that the DICE-distilled student preserves the complete DFT relative-energy ordering, whereas the directly fine-tuned student changes the ordering of some intermediate-energy configurations (Fig. 5e; Supplementary Fig. 7). These comparisons show that expanding the student training set with teacher labels makes more effective use of the same DFT references, yielding more reliable compact potentials than direct fine-tuning.\par

\FloatBarrier

\subsection*{DICE potentials reproduce material properties and accelerate exploration}

\noindent We further assess the physical-property predictions of the three final DICE student potentials across different structures, compositions, temperatures and pressures against DFT calculations and experiments, and evaluate exploration efficiency using MD benchmarks and actual development records.\par

The same Al\textsubscript{2}O\textsubscript{3} student potential describes crystalline, liquid and solid--liquid coexistence behaviour, with complementary crystal-relaxation and phonon comparisons in Supplementary Figs. 8,9. It reproduces the energy--volume curves and relative energies of five polymorphs, with fitted equilibrium volumes differing from DFT by less than 0.4\% (Fig. 6a). At 2400 K, simulated X-ray and neutron total correlation functions of liquid Al\textsubscript{2}O\textsubscript{3} reproduce the main experimental peak positions and subsequent oscillations \cite{dice_ref_18} (Fig. 6b). In solid--liquid coexistence simulations, the mean interface velocity approaches zero near 2340 K, giving a melting-temperature estimate close to the experimental reference of 2327 \ensuremath{\pm} 6 K \cite{dice_ref_19} (Supplementary Fig. 10).\par

\begin{figure}[!htbp]\centering
\includegraphics[width=180.00mm,height=180.00mm,keepaspectratio]{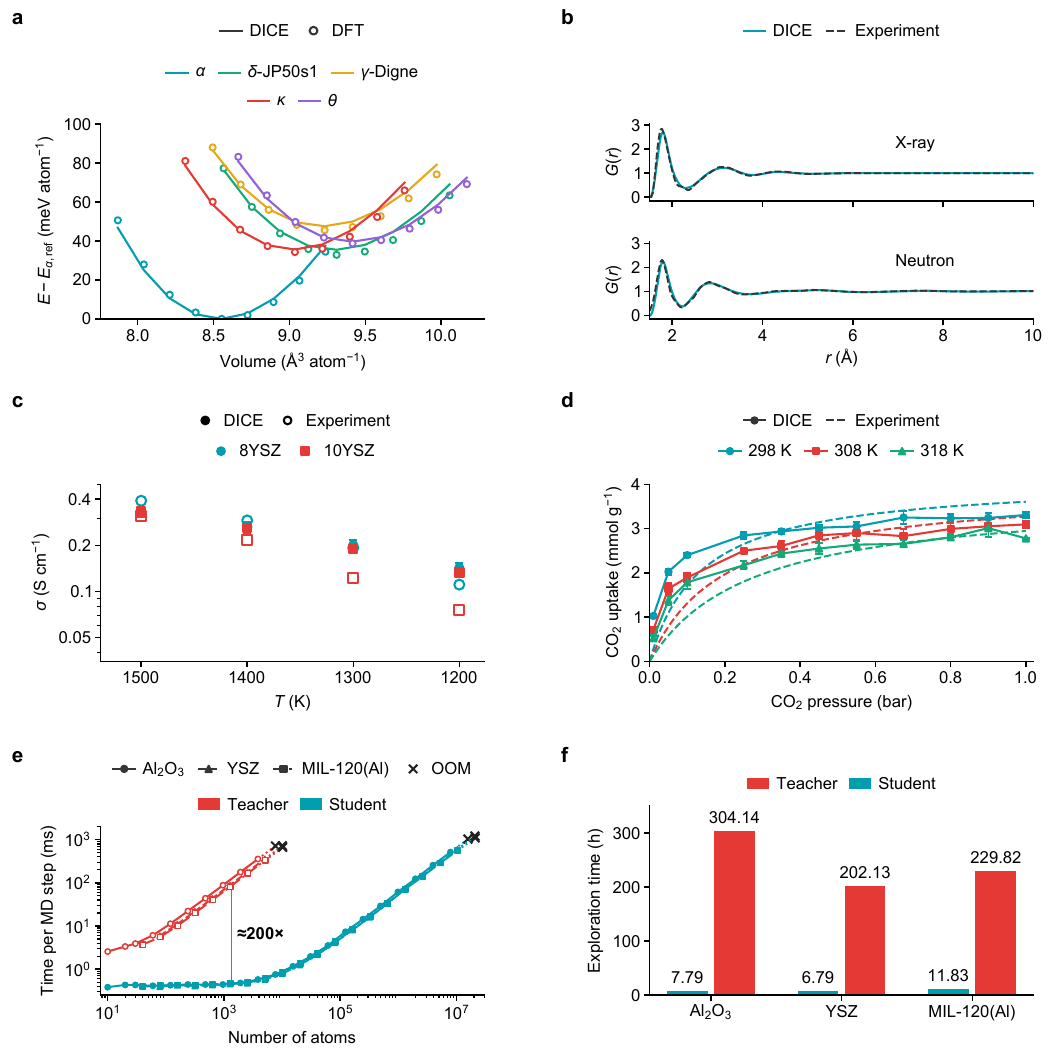}
\par\vspace{2mm}{\fontsize{10}{12}\selectfont\raggedright\textbf{Fig. 6. Material properties and configuration-exploration efficiency of DICE compact potentials.} a, Equations of state for five Al\textsubscript{2}O\textsubscript{3} polymorphs, with each method's $\alpha$-phase energy at the reference volume as zero. b, Liquid Al\textsubscript{2}O\textsubscript{3} X-ray and neutron total correlation functions at 2400 K versus experiment \cite{dice_ref_18}. c, 8YSZ and 10YSZ conductivities versus experimental DC data at nearby compositions \cite{dice_ref_21} (open points). d, CO\textsubscript{2} adsorption in flexible MIL-120(Al) versus experimental Langmuir fits \cite{dice_ref_20} (dashed curves). b--d show means and sample SDs across three trajectories, four independent structural preparations and three stochastic repeats, respectively. e, MD-step times versus atom count for the DPA4-Neo teacher and DPA4C-Nano student on one NVIDIA RTX 4090 GPU; points and bars show medians and interquartile ranges across five post-warm-up timing blocks, and crosses mark first tested out-of-memory sizes. f, Total recorded student exploration times per system versus corresponding teacher estimates from measured execution rates and task overheads for matched development workloads.\par}
\end{figure}

The YSZ student potential predicts a non-monotonic dependence of oxygen tracer diffusivity on yttria content, with higher values at intermediate dopant concentrations followed by a decrease with further doping, in qualitative agreement with the experimental composition trend \cite{dice_ref_22} (Supplementary Fig. 11). Conductivities estimated through the Nernst--Einstein relation for 8YSZ (nominally 8 mol\% Y\textsubscript{2}O\textsubscript{3}) and 10YSZ reproduce the temperature dependence and order of magnitude of experimental DC conductivities at nearby compositions over 1200--1500 K \cite{dice_ref_21} (Fig. 6c).\par

For MIL-120(Al), we use the final DICE student potential to optimise nine hydroxyl-rearrangement pathways with the climbing-image nudged elastic band (CI-NEB) method, keeping the DFT endpoints and cells fixed. The resulting forward and reverse barriers agree well with DFT (Supplementary Fig. 12). This extends energy comparisons on matched DFT configurations to potential-optimised pathways. The same student potential is also used to simulate CO\textsubscript{2} adsorption in flexible MIL-120(Al) at 298--318 K and 0.01--1 bar. Uptake increases with pressure and decreases with temperature, consistent with the overall trends of published experimental Langmuir fits \cite{dice_ref_20} (Fig. 6d).\par

The compact student architecture also offers substantial advantages in computational speed and accessible system size. Across all three systems' common size ranges, single-trajectory NVT benchmarks show substantially lower time per MD step for the student than the teacher (Fig. 6e). Student step times remain near 0.4 ms for small cells before increasing with size. On the same GPU, the student completes the Al\textsubscript{2}O\textsubscript{3} benchmark at approximately 7.9 million atoms, whereas the teacher's largest successful system contains 3840 atoms.\par

This speed advantage also translates into shorter configuration-exploration times during development. For matched exploration workloads contributing to the final training sets, the per-system totals of recorded student time are reduced to approximately 2.6--5.1\% of the corresponding teacher estimates from measured execution rates and task overheads (Fig. 6f; Supplementary Note 1).\par

Despite substantial differences in chemical composition, local bonding environments and target properties, DICE yields compact student potentials that reliably reproduce key DFT results and experimental trends across all three systems while substantially reducing configuration-exploration times. These results demonstrate the framework's applicability across materials systems and its ability to combine reliable potentials with efficient exploration.\par

\FloatBarrier

\section*{Discussion}

\noindent DICE changes the division of work between teachers and students during potential development, extending compact-model acceleration from final deployment to training-data acquisition. Compared with two-stage distillation that adapts the teacher before training students, DICE assigns exploration to a continually learning student and alternates teacher adaptation with student training within the same workflow. Applications to Al\textsubscript{2}O\textsubscript{3}, YSZ and MIL-120(Al) demonstrate lower configuration-exploration costs alongside reliable compact student potentials.\par

This workflow uses both new DFT information and knowledge already learned by the teacher. Teacher updates improve predictions on unqueried configurations, allowing the same DFT references to contribute to student training through more teacher labels. Existing reliable teacher predictions offer a complementary source of training data, improving student accuracy without additional target DFT references. Support-guided routing selects configurations for labelling by the current teacher and supplies low-support candidates for DFT labelling for subsequent teacher updates, making both supervision sources available during continuing exploration. Single-teacher support assessment avoids committee training and evaluation costs and is well suited to continuing teacher updates.\par

Support-guided routing depends on the teacher representations' ability to distinguish relevant chemical environments and on the reference data used for comparison. Here, empirical support is defined relative to teacher training data, giving thresholds on $p$ a clear statistical interpretation relative to the reference distribution; support is reassessed after teacher updates. The threshold separates configurations for teacher labelling from candidates for DFT labelling. We evaluate the resulting supervision against students trained entirely on DFT labels for matched training geometries, assessing whether expanding the training set maintains student accuracy.\par

Statistics of representations from pretraining data, combined with complementary reliability-assessment methods \cite{dice_ref_11,dice_ref_23}, could identify more teacher predictions usable for student training. With appropriate training, labelling and representation interfaces, this workflow could extend to other teacher--student combinations. Overall, DICE combines fast student exploration, selective DFT feedback and continuing teacher adaptation in a compact-potential development workflow integrating configuration-exploration efficiency, DFT data efficiency and supervision quality for complex materials.\par

\section*{Methods}

\subsection*{DICE workflow and support assessment}

\noindent Student-explored candidates undergo pair-distance checks and joint DIRECT preselection \cite{dice_ref_08,liu2026dpeva} before support assessment. After support assessment, DIRECT selects up to the configured quota of high-support configurations per input system and cycle for teacher labelling and addition to the student training set. The support estimator is rebuilt after each teacher update for the next iteration.\par

Support is calculated from 96-dimensional atomic representations extracted from the teacher's fitting network. Representations are standardised separately by element and reduced using principal component analysis (PCA), retaining 99\% of the variance with at most 32 components. An eight-component full-covariance Gaussian mixture model (GMM) is fitted separately for each element using scikit-learn \cite{dice_ref_26,pedregosa2011sklearn}. For atom $i$, the negative log fitted density defines $\ell_i$, and the atomic support score $\pi_i$ is calculated by empirical ranking against reference atoms of the same element:\par

\begin{equation}\ell_i=-\ln\rho_{Z_i}(\mathbf z_i),\qquad
\pi_i=
\frac{1+\sum_{j\in R_{Z_i}}\mathbf 1(\ell_j\geq\ell_i)}
{|R_{Z_i}|+1}.
\tag{1}\end{equation}

Here, $\mathbf z_i$ is the reduced representation of atom $i$, $\rho_{Z_i}$ is the fitted probability density for its element, and $R_{Z_i}$ is the set of reference atoms of that element. The indicator $\mathbf 1(A)$ equals one when condition $A$ holds and zero otherwise.\par

For a configuration containing $N$ atoms, the $k=\lceil0.05N\rceil$ atoms with the lowest support are selected. Their mean negative log support defines a nonconformity score $s$, which is converted to the configuration support score $p$ by empirical ranking against reference configurations:\par

\begin{equation}s=-\frac{1}{k}\sum_{i\in I}\ln\pi_i,\qquad
p=
\frac{1+\sum_{r\in R}\mathbf 1(s_r\geq s)}
{|R|+1}.
\tag{2}\end{equation}

Here, $I$ indexes the selected atoms and $R$ is the teacher training set. Routing uses a support threshold of 0.1: configurations with $p>0.1$ are classified as suitable for reliable teacher labelling, whereas those with $p\leq0.1$ require further feedback and enter the DFT-labelling candidate pool.\par

\subsection*{Model training and DFT calculations}

\noindent Teachers and students are fine-tuned from OMat24-pretrained DPA4-Neo and DPA4C-Nano, respectively \cite{dice_ref_14,dice_ref_15,dice_ref_16}, using DeePMD-kit \cite{dice_ref_25} with a 6 Å cutoff. Teachers use low-rank adaptation (LoRA) \cite{dice_ref_27}, with both rank and scaling factor set to 16. Training uses the HybridMuon optimiser and mean absolute error (MAE) losses for energy, forces and virial.\par

DICE-distilled students and directly fine-tuned students start from the same pretrained DPA4C-Nano model and use identical fine-tuning settings. DICE-distilled students use training datasets expanded through teacher labelling, whereas directly fine-tuned students use the target DFT references acquired during development. Final student potentials for all three systems are fine-tuned for 500,000 steps. Final teachers for Al\textsubscript{2}O\textsubscript{3}, YSZ and MIL-120(Al) are fine-tuned for 25,000, 40,000 and 40,000 steps, respectively.\par

Teachers in the controlled experiments are fine-tuned for 5,000 steps.\par

First-principles calculations are performed using the Vienna Ab initio Simulation Package (VASP) \cite{dice_ref_28}, with electron--ion interactions treated using the projector augmented-wave (PAW) method \cite{dice_ref_29,paw1999}. Al\textsubscript{2}O\textsubscript{3} and YSZ use the PBEsol exchange--correlation functional \cite{dice_ref_30}, with plane-wave cutoffs of 600 and 800 eV, respectively. Calculations for the 40-atom MIL-120(Al) framework cell (P40) use PBE with the D3 dispersion correction \cite{dice_ref_31,dice_ref_32} and a 1,000 eV cutoff.\par

Al\textsubscript{2}O\textsubscript{3} uses \ensuremath{\Gamma}-centred $k$-point meshes generated with KSPACING = 0.15 Å\textsuperscript{\ensuremath{-}1}. YSZ uses a $2\times2\times2$ Monkhorst--Pack mesh \cite{monkhorst1976}, and MIL-120(Al) P40 uses a \ensuremath{\Gamma}-centred $5\times5\times6$ mesh.\par

\subsection*{Model evaluation and structural analysis}

\noindent Controlled teacher-feedback experiments use the DFT-labelled $\gamma$-Smr\v{c}ok and liquid--amorphous Al\textsubscript{2}O\textsubscript{3} datasets of Zhang \textit{et al.} \cite{dice_ref_33,dice_ref_34}, containing 516 and 423 configurations, respectively. A common initial teacher $T_0$ is fine-tuned from the pretrained DPA4-Neo model on ten $\gamma$-Smr\v{c}ok references. In each of five configuration-selection repeats, sixty configurations are drawn randomly from the remaining $\gamma$-Smr\v{c}ok pool or the liquid--amorphous pool and combined with the initial ten to define seventy common student-training geometries.\par

The $T_0$ and DFT-70 routes use teacher and DFT labels, respectively, on all seventy geometries. Ten of the sixty additions are selected randomly for DFT feedback. Hybrid replaces their labels with DFT labels while retaining $T_0$ labels on the other sixty geometries, including the initial ten. The teacher-update route fine-tunes $T_R$ from the same pretrained DPA4-Neo model on the initial ten and these ten new DFT references, then uses $T_R$ to relabel all seventy geometries. DFT-20 trains the student only on the twenty DFT references. All routes start from the same pretrained DPA4C-Nano model, use identical fine-tuning settings and share the evaluation set after excluding student-training configurations. Teacher errors are evaluated separately on all seventy training geometries, the fifty additions excluded from teacher adaptation, and the held-out set.\par

For QBC screening, the committee comprises $T_0$ and three additional teachers fine-tuned from the same pretrained DPA4-Neo model on the same initial ten $\gamma$-Smr\v{c}ok configurations for 5,000 steps. Model and data-sampling random seeds differ between members. Configuration disagreement $\sigma$ is the maximum, across atoms, of the root-mean-square deviation of the four teacher force vectors from their mean. Configurations with $\sigma$ at or below the tested threshold of 0.03, 0.05 or 0.08 eV Å\textsuperscript{\ensuremath{-}1} are accepted for teacher labelling; the remainder require further feedback. Each student comparison combines the initial ten geometries with fifty randomly selected accepted configurations, using either $T_0$ or DFT labels on the same sixty geometries.\par

Initial-teacher GMM comparisons combine the ten original references with fifty accepted $\gamma$-Smr\v{c}ok configurations or all five accepted liquid--amorphous configurations, giving sixty and fifteen student-training geometries, respectively. In each pool, ten selected low-support DFT references supplement the initial ten to fine-tune $T_1$ from the same pretrained DPA4-Neo model. Support is reassessed, fifty additional accepted configurations are selected, and $T_1$ relabels the expanded student set, including the ten new DFT references. The resulting sets contain 120 and seventy-five configurations, respectively. Newly selected configurations are drawn randomly without replacement and exclude those already in student training. Students trained on teacher labels are compared with students trained entirely on DFT labels using identical training geometries.\par

For routing diagnostics, a configuration is positive when its teacher force-component RMSE exceeds 0.05 eV Å\textsuperscript{\ensuremath{-}1}. A positive configuration accepted for teacher labelling is a false negative (FN), whereas an accepted configuration at or below the error threshold is a true negative (TN). The fraction of high-error configurations among those accepted for teacher labelling is FN/(TN+FN).\par

Independent test sets for the final student potentials contain 200 Al\textsubscript{2}O\textsubscript{3}, 330 YSZ and 300 MIL-120(Al) configurations without overlap with training data. All models are evaluated on the same test geometries. Energy errors use per-atom residuals without additional energy offsets, and force-component root-mean-square error (RMSE) pools all Cartesian components. The per-atom force-error magnitude is the Euclidean norm of the difference between predicted and DFT forces. Upper-tail force errors are calculated by ranking all atom--configuration observations by per-atom force-error magnitude and averaging over the specified highest-error fraction.\par

Controlled comparisons use five repeats: five training seeds on fixed geometries for the initial 15-configuration GMM comparison, and five configuration-selection repeats for the other comparisons. Results are summarised as the mean and sample standard deviation. Differences in student upper-tail force errors relative to the matched control trained entirely on DFT labels are calculated within each repeat before averaging across repeats. Within each repeat, the relative P99 difference is calculated as the difference between the P99 values for students trained on teacher and DFT labels, divided by the latter and multiplied by 100.\par

Paired comparisons of YSZ per-atom force-error magnitudes use 42 source-structure clusters as the bootstrap resampling units. Observations within each cluster are sampled together, and identical cluster draws are used for both models to preserve the correspondence between test structures. Element-resolved high-error fractions are recalculated for 1,000 bootstrap resamples, with the 2.5th and 97.5th percentiles defining the 95\% confidence intervals.\par

For the comparisons in Fig. 4a,b, the oxygen--oxygen partial structure factor is evaluated at reciprocal-lattice vectors $\mathbf G$ of each periodic cell:\par

\begin{equation}S_{\mathrm{OO}}(\mathbf G)=\frac{1}{N_{\mathrm O}}\left|\sum_{j\in\mathrm O}\exp(i\mathbf G\cdot\mathbf r_j)\right|^2.
\tag{3}\end{equation}

Here, $N_{\mathrm O}$ is the number of oxygen atoms and $\mathbf r_j$ an oxygen position. Radial averaging uses a Gaussian kernel of standard deviation 0.08 Å\textsuperscript{\ensuremath{-}1} over 1--10 Å\textsuperscript{\ensuremath{-}1} at 0.005 Å\textsuperscript{\ensuremath{-}1} spacing, with equal configuration weights within each dataset. Oxygen-network bond-orientational order uses the neighbour-averaged $\bar q_6$ parameter \cite{lechner2008} with the twelve nearest oxygen neighbours; $\langle\bar q_6\rangle$ is its mean over oxygen atoms in each configuration. The normalised reciprocal-space peak intensity is\par

\begin{equation}I_{\mathrm O}=\max_{1.5\leq|\mathbf G|<5\ \text{\AA}^{-1}}\frac{S_{\mathrm{OO}}(\mathbf G)}{N_{\mathrm O}}.
\tag{4}\end{equation}

For Al- and O-centred environments separately, the fine-tuned teacher representations of the teacher training set and liquid--amorphous test set undergo joint standardisation, PCA reduction and t-SNE projection \cite{dice_ref_35}. Within radius 8 t-SNE units of environment $i$, neighbours from each source are counted and divided by that source's total number of environments, excluding $i$ from both the neighbour count and its source total. Let $u_i$ be the teacher contribution divided by the sum of both source-normalised contributions. Neighbourhood mixing is\par

\begin{equation}M_i=4u_i(1-u_i).
\tag{5}\end{equation}

Values of $M_i=0$ and $M_i=1$ indicate single-source neighbourhoods and equal source-normalised contributions, respectively.\par

Local Al--O environments are identified by periodic coordination analysis in the Atomic Simulation Environment (ASE) \cite{larsen2017ase}, including a central Al and all O neighbours within 2.8 Å. Examples are selected from the eighty test configurations with both oxygen-network order metrics below the teacher-training minima. Within each Al coordination class, all teacher--test environment pairs are compared after superposing Al centres; O-position RMSD is minimised over O-site permutations and proper rotations. The minimum-RMSD pair is selected and rendered using VESTA \cite{momma2011vesta}.\par

\subsection*{Atomistic simulations and physical-property calculations}

\noindent Molecular dynamics simulations use LAMMPS \cite{dice_ref_37}. Crystalline Al\textsubscript{2}O\textsubscript{3} equations of state are obtained by internal-coordinate relaxation of five polymorphs at different volumes and Birch--Murnaghan fitting \cite{dice_ref_38}. Phonon dispersions are calculated using Phonopy \cite{dice_ref_39,dice_ref_40}. The $\alpha$-Al\textsubscript{2}O\textsubscript{3} phonon comparison between directly fine-tuned and DICE-distilled students uses a 120-atom cell and seven displacement configurations supplied by NIMS \cite{dice_ref_41}. Model phonon modes are matched to DFT modes by eigenvector overlap. Frequency-error statistics include modes with DFT frequencies above 0.5 THz, while full dispersion plots retain all branches.\par

Liquid Al\textsubscript{2}O\textsubscript{3} structure is evaluated using three 1,920-atom NPT trajectories simulated for 200 ps at 2400 K and zero pressure. X-ray and neutron total correlation functions are calculated and compared with experiment \cite{dice_ref_18}. Solid--liquid coexistence simulations use a 23,040-atom system at 2330--2350 K and zero pressure in NPT, with three 1 ns trajectories with different initial velocities per temperature. The first 100 ps of each coexistence trajectory are discarded before interface-velocity analysis.\par

YSZ simulations cover five compositions from 5.8 to 14.3 mol\% Y\textsubscript{2}O\textsubscript{3} in cells containing 256 cations. Four independent structural preparations per composition are simulated for 5 ns in NVT at 1200--1600 K. Oxygen tracer diffusivity $D_{\mathrm O}^{*}$ is obtained by linear fitting of the oxygen mean-squared displacement (MSD) over lag times of 100--200 ps after removing the net translation of the cation sublattice. Conductivity is estimated using the Nernst--Einstein relation \cite{dice_ref_43}:\par

\begin{equation}\sigma_{\mathrm{NE}}
=
\frac{N_{\mathrm O}(2e)^2D_{\mathrm O}^{*}}
{V k_{\mathrm B}T}.
\tag{6}\end{equation}

Here, $N_{\mathrm O}$ is the number of oxygen ions and $V$ is the cell volume. Calculated values are compared with digitised experimental DC conductivity at nearby compositions \cite{dice_ref_21}, using interpolation in the original plotting coordinates without temperature extrapolation.\par

Relative energies of the six MIL-120(Al) framework configurations with different hydroxyl orientations are obtained from the minima of equation-of-state fits to their energy--volume curves \cite{dice_ref_17}. Nine hydroxyl-rearrangement pathways are evaluated on common DFT geometries and on climbing-image nudged elastic band (CI-NEB) paths optimised using the compact potential, with the latter retaining fixed DFT endpoints and cells \cite{dice_ref_42,henkelman2000tangent,bitzek2006fire}. DICE CI-NEB optimisation uses a final climbing-image force threshold of 0.03 eV Å\textsuperscript{\ensuremath{-}1}.\par

CO\textsubscript{2} adsorption uses a flexible 160-atom MIL-120(Al) framework and fixed-cell hybrid grand-canonical Monte Carlo--molecular dynamics (GCMC--MD) simulations at 298, 308 and 318 K and 0.01--1 bar. The MD timestep is 0.5 fs, with one randomly selected CO\textsubscript{2} insertion or deletion attempt every five MD steps. Three 500 ps stochastic repeats are performed for each temperature--pressure condition. Adsorption is averaged over 100--500 ps and compared with published experimental Langmuir fits \cite{dice_ref_20}.\par

Variability in liquid total correlation functions and adsorption results is reported as the sample standard deviation across three trajectories and three stochastic repeats, respectively. Conductivity variability is reported as the sample standard deviation across four independently prepared structures.\par

Supplementary Note 5 gives the auxiliary phonon, coexistence, diffusivity, thermal-expansion (Supplementary Fig. 13) and pathway procedures.\par

\subsection*{Computational performance and exploration costs}

\noindent Computational efficiency is benchmarked using the pretrained DPA4-Neo teacher and DPA4C-Nano student on a single NVIDIA RTX 4090 GPU. Single-trajectory NVT simulations at 300 K use periodic $\alpha$-Al\textsubscript{2}O\textsubscript{3}, 8YSZ and MIL-120(Al) empty-framework cells of different sizes. Time per step includes model computation and neighbour handling during MD integration and is summarised by the median and interquartile range across five consecutive post-warm-up timing intervals. System size is increased until the first out-of-memory event.\par

Configuration-exploration costs are calculated from the actual exploration records associated with the final training datasets. Student time sums recorded elapsed times for the accepted exploration task blocks. Teacher time combines measured aggregate execution rates and task overheads for the same workload, retaining the corresponding teacher, native configuration-generation operations and output settings. Estimated teacher time for system $s$ is:\par

\begin{equation}\widehat T_{\mathrm{teacher},s}
=
\sum_{b\in s}\sum_m
\left(\frac{W_{b,m}}{r^{\mathrm{teacher}}_{b,m}}+H^{\mathrm{teacher}}_{b,m}\right).
\tag{7}\end{equation}

Here, $W_{b,m}$ is the observed work in block $b$ and mode $m$, $r^{\mathrm{teacher}}_{b,m}$ is the measured aggregate teacher execution rate, and $H^{\mathrm{teacher}}_{b,m}$ includes startup, exit, MD-subprocess overheads, preparation and result validation.\par

Detailed timing measurements and workload accounting are provided in Supplementary Note 1.\par

\section*{Data availability}

\noindent The training and test datasets, trained models and numerical source data supporting this study are available at \url{https://doi.org/10.5281/zenodo.24456706}\cite{zenodo2026dice}.\par

\section*{Code availability}

\noindent The DICE implementation and the analysis and figure-reproduction code are available at \url{https://github.com/NanaKk0/DICE} and \url{https://github.com/NanaKk0/DICE-paper}, respectively. Archived versions are available at \url{https://doi.org/10.5281/zenodo.24456706}\cite{zenodo2026dice}.\par

\renewcommand{\refname}{References}

\begingroup\setstretch{1.05}\fontsize{11}{13}\selectfont\endgroup

\section*{Acknowledgements}

\noindent Calculations were performed using the IBM Blade cluster at the High Performance Computing Center (HPCC) of Nanjing University. We also acknowledge computational resources provided by the Open Source Supercomputing Center of S-A-I. We thank Tiancheng Li (Peking University, and AI for Science Institute) for helpful discussions.\par

\section*{Funding}

\noindent This work was supported by the National Key R\&D Program of China (No. 2025YFE0107600) and the Fundamental Research Funds for the Central Universities (0211/14380232).\par

\section*{Author contributions}

\noindent R.D. developed the support-guided routing method and DICE software, designed and conducted the computational experiments, analysed the data and drafted the manuscript. H.H. conceived the overall DICE framework and contributed to methodological development. Y.H. contributed data and advised on DFT calculations. G.Z. supervised the research and guided manuscript preparation.\par

\section*{Competing interests}

\noindent The authors declare no competing interests.\par

%% file: supplement-body.tex
\hypersetup{pdfauthor={Ruiqi Dong; Haotian Huang; Yubin Hu; Guixiang Zeng}}

{\raggedright\noindent\fontsize{16}{19}\selectfont\bfseries Supplementary Information\par}\vspace{4mm}

{\raggedright\noindent\fontsize{13}{16}\selectfont\bfseries Accelerating interatomic potential development through student-driven exploration and reliability-aware distillation\par}\vspace{4mm}

\noindent Ruiqi Dong\textsuperscript{1}, Haotian Huang\textsuperscript{1,2}, Yubin Hu\textsuperscript{1}, Guixiang Zeng\textsuperscript{1*}\par

\noindent\textsuperscript{1}Kuang Yaming Honors School, Nanjing University, Nanjing 210023, China\par

\noindent\textsuperscript{2}School of Chemistry, Nanjing University, Nanjing 210023, China\par

\noindent\textsuperscript{*}Correspondence: gxzeng@nju.edu.cn\par\vspace{4mm}

\section*{Contents}

\noindent Supplementary Note 1 | Development data and exploration costs\par\smallskip

\noindent Supplementary Note 2 | Controlled teacher-feedback experiments\par\smallskip

\noindent Supplementary Note 3 | QBC and support-guided routing\par\smallskip

\noindent Supplementary Note 4 | DICE distillation versus direct fine-tuning\par\smallskip

\noindent Supplementary Note 5 | Validation of physical properties\par\smallskip

\clearpage

\section*{Supplementary Note 1 | Development data and exploration costs}

\subsection*{S1.1 Configuration-exploration cost accounting}

\noindent Teacher timings are measured on one RTX 4090 GPU for each combination of configuration-generation method, chemical system, atom-count range and output cadence, retaining the corresponding teacher and native operations. Concurrency maximises aggregate throughput, up to the number of tasks available in each block. Three independent fresh-process runs provide execution rates and startup/exit overheads; medians are used.\par

Startup includes model loading and the first GPU evaluation. Startup counts follow concurrent task batches, with persistent FIRE workers initialised once per block and GCMC--MD subprocess overheads counted per MD segment. Preparation and result-validation times are recovered from the corresponding historical block logs.\par

Work is counted as MD steps, relaxation force evaluations or Widom insertion proposals, including model energy evaluations. Hybrid GCMC--MD counts Monte Carlo attempts plus five times the MD-step count. Recorded student times are allocated among modes using native-timing weights.\par

\clearpage

\noindent\begin{minipage}{\linewidth}\setstretch{1}\fontsize{9}{10.8}\selectfont
\textbf{Supplementary Table 1 | Al\textsubscript{2}O\textsubscript{3} development stages contributing to the final training set.}\par\vspace{3mm}
\setlength{\tabcolsep}{3pt}\renewcommand{\arraystretch}{1.12}
\begin{tabular}{@{}>{\raggedright\arraybackslash}p{\dimexpr 0.180000\linewidth-5.4000pt\relax}>{\raggedright\arraybackslash}p{\dimexpr 0.210000\linewidth-6.3000pt\relax}>{\raggedright\arraybackslash}p{\dimexpr 0.270000\linewidth-8.1000pt\relax}>{\raggedright\arraybackslash}p{\dimexpr 0.160000\linewidth-4.8000pt\relax}>{\raggedright\arraybackslash}p{\dimexpr 0.090000\linewidth-2.7000pt\relax}>{\raggedright\arraybackslash}p{\dimexpr 0.090000\linewidth-2.7000pt\relax}@{}}\toprule
\textbf{Stage} & \textbf{Initial con\-figu\-ra\-tions} & \textbf{Exploration conditions} & \textbf{Run length} & \textbf{New teacher-labelled con\-figu\-ra\-tions} & \textbf{New DFT reference con\-figu\-ra\-tions} \\\midrule
Initial crystal references\newline (iteration 0) & Crystal structures and strain/\allowbreak displacement variants & DFT references & --- & 0 & 20 \\
Low-temperature crystal exploration\newline (iteration 0--2) & Crystal seeds and strain/\allowbreak displacement variants, 80/\allowbreak 160 atoms & NVT; 100, 300 K & 0.1 ps & 240 & 40 \\
Extended crystal exploration\newline (iteration 2--6) & Crystal seeds and strain/\allowbreak displacement variants; 30/\allowbreak 80/\allowbreak 160 atoms & NVT; 500, 1000, 1500, 2000 K & 0.1 ps & 560 & 80 \\
Crystal-heating exploration\newline (iteration 6 r2--r6) & Crystal seeds, 160 atoms & NVT ramps: 300 \ensuremath{\rightarrow} 2600/\allowbreak 3200/\allowbreak 3800/\allowbreak 4200 K; hold at target & 50 ps & 230 & 0 \\
Low-density melt--cool exploration\newline (iteration 6 r7--r9) & Low-density seeds, 160 atoms & 300 \ensuremath{\rightarrow} 3000 K; densification to 3.5 then 4.1 g cm\textsuperscript{\ensuremath{-}3}; slow cooling or quench/\allowbreak anneal & 350/\allowbreak 360 ps & 150 & 0 \\
Crystal temperature/\allowbreak pressure expansion\newline (iteration 6 r10--r12) & Crystal seeds, 160 atoms & NPT; 100--2000 K; 0, 5, 20 GPa & 20/\allowbreak 30/\allowbreak 40 ps & 240 & 0 \\
Thermoelastic crystal expansion\newline (iteration 6 r13--r15) & Strained $\alpha$ seeds, 80 atoms & NVT; 300, 600, 900, 1200, 1500, 1800, 2000 K & 20/\allowbreak 30/\allowbreak 40 ps & 240 & 0 \\
Long-wavelength displacement relaxation\newline (iteration 6 r16--r17) & Displaced $\alpha$ seeds, 120 atoms & Fixed-cell FIRE \cite{si_bitzek2006fire}; force criterion 0.02 eV Å\textsuperscript{\ensuremath{-}1} & \ensuremath{\leq}400 optimiser steps & 160 & 0 \\
Displaced-crystal dynamics\newline (iteration 6 r18--r21) & Unstrained/\allowbreak strained $\alpha$ seeds, 120 atoms & NVT; 300, 500, 700 K; also 1100 K for unstrained seeds & 50 ps & 320 & 0 \\
Longitudinal/\allowbreak transverse relaxation\newline (iteration 6 r22--r23) & Displaced $\alpha$ seeds, 240/\allowbreak 360 atoms & Fixed-cell FIRE; force criterion 0.01 eV Å\textsuperscript{\ensuremath{-}1} & \ensuremath{\leq}400 optimiser steps & 100 & 0 \\
Strain-coupled crystal dynamics\newline (iteration 6 r24--r27) & Pure/\allowbreak coupled-strain $\alpha$ seeds, 240 atoms & NVT; 100, 300, 500 K & 50 ps & 200 & 0 \\
Total &  &  &  & 2,440 & 140 \\
\bottomrule\end{tabular}
\par\vspace{2mm}In Supplementary Tables 1--3, teacher-labelled counts refer to DIRECT-selected high-support additions retained in the final student training sets; DFT references are counted separately. Run lengths are specified per task. In iteration 6, 7,949 of 8,000 prescreened candidates in r2--r9 satisfied $p>0.1$; DIRECT retained forty configurations per cycle in r2--r3 and fifty in r4--r9, yielding 380 additions. Al\textsubscript{2}O\textsubscript{3} crystal seed pools include lattice-strained structures and random atomic displacements with standard deviations of 0.02, 0.05 and 0.10 Å. The final Al\textsubscript{2}O\textsubscript{3} student training set contains 2,580 configurations.
\end{minipage}\par\medskip

\clearpage

\noindent\begin{minipage}{\linewidth}\setstretch{1}\fontsize{9}{10.8}\selectfont
\textbf{Supplementary Table 2 | YSZ development stages contributing to the final training set.}\par\vspace{3mm}
\setlength{\tabcolsep}{3pt}\renewcommand{\arraystretch}{1.12}
\begin{tabular}{@{}>{\raggedright\arraybackslash}p{\dimexpr 0.180000\linewidth-5.4000pt\relax}>{\raggedright\arraybackslash}p{\dimexpr 0.210000\linewidth-6.3000pt\relax}>{\raggedright\arraybackslash}p{\dimexpr 0.270000\linewidth-8.1000pt\relax}>{\raggedright\arraybackslash}p{\dimexpr 0.160000\linewidth-4.8000pt\relax}>{\raggedright\arraybackslash}p{\dimexpr 0.090000\linewidth-2.7000pt\relax}>{\raggedright\arraybackslash}p{\dimexpr 0.090000\linewidth-2.7000pt\relax}@{}}\toprule
\textbf{Stage} & \textbf{Initial con\-figu\-ra\-tions} & \textbf{Exploration conditions} & \textbf{Run length} & \textbf{New teacher-labelled con\-figu\-ra\-tions} & \textbf{New DFT reference con\-figu\-ra\-tions} \\\midrule
Initial bulk references\newline (iteration 0) & YSZ bulk reference structures & DFT references & --- & 0 & 15 \\
Initial temperature exploration\newline (iteration 0--2) & YSZ cells, 64/\allowbreak 79/\allowbreak 92 atoms & NPT; 300, 1200, 1800 K; 1 bar & 20 ps & 180 & 45 \\
Auxiliary dimer references\newline (iteration 3) & Six pair types; three separations each & DFT references & --- & 0 & 18 \\
Extended temperature exploration\newline (iteration 3--4) & YSZ cells, 64/\allowbreak 79/\allowbreak 92 atoms & NPT; 300, 1200, 1800 K; 1 bar & 50/\allowbreak 75/\allowbreak 100 ps & 240 & 30 \\
Longer NPT trajectories\newline (iteration 5 r0--r3) & YSZ cells, 64/\allowbreak 79/\allowbreak 92 atoms & NPT; 300, 1200, 1800 K; 1 bar & 100/\allowbreak 150/\allowbreak 200 ps & 480 & 0 \\
Intermediate-temperature expansion\newline (iteration 5 r4--r7) & YSZ cells, 64/\allowbreak 79/\allowbreak 92 atoms & NPT; 600, 900, 1500 K; 1 bar & 50/\allowbreak 100/\allowbreak 150/\allowbreak 200 ps & 360 & 0 \\
Fixed-volume expansion\newline (iteration 5 r8--r11) & YSZ cells; volumes 0.97/\allowbreak 1.03 of prepared cells & NVT; 600, 1200 K & 50/\allowbreak 100/\allowbreak 150/\allowbreak 200 ps & 480 & 0 \\
Temperature-programmed exploration\newline (iteration 5 r12--r19) & YSZ cells, 64/\allowbreak 79/\allowbreak 92 atoms & NPT; 1800 \ensuremath{\rightarrow} 1200 \ensuremath{\rightarrow} 300 K; 1 bar & 500 ps & 720 & 0 \\
Short-range dimer extension\newline (iteration 6) & Six pair types; two shorter separations each & DFT references & --- & 0 & 12 \\
High-temperature expansion\newline (iteration 6--7) & YSZ cells, 64/\allowbreak 79/\allowbreak 92 atoms & NPT; 2100, 2400, 2700 K; 1 bar & 25/\allowbreak 50/\allowbreak 75/\allowbreak 100/\allowbreak 150/\allowbreak 200 ps & 540 & 15 \\
Compressed-cell feedback\newline (iteration 7--9) & YSZ cells at 0.80 prepared volume & NVT; 2700 K & 10/\allowbreak 20/\allowbreak 40 ps & 447 & 45 \\
Longer compressed-cell dynamics\newline (iteration 10 r0--r2) & YSZ cells at 0.80 prepared volume & NVT; 2700 K & 50/\allowbreak 100/\allowbreak 150 ps & 270 & 15 \\
Final compressed-cell expansion\newline (iteration 11 r0--r3) & YSZ cells at 0.80 prepared volume & NVT; 2700 K & 200 ps & 360 & 0 \\
Total &  &  &  & 4,077 & 195 \\
\bottomrule\end{tabular}
\par\vspace{2mm}Final student training set: 4,272 configurations.
\end{minipage}\par\medskip

\clearpage

\noindent\begin{minipage}{\linewidth}\setstretch{1}\fontsize{9}{10.8}\selectfont
\textbf{Supplementary Table 3 | MIL-120(Al) development stages contributing to the final training set.}\par\vspace{3mm}
\setlength{\tabcolsep}{3pt}\renewcommand{\arraystretch}{1.12}
\begin{tabular}{@{}>{\raggedright\arraybackslash}p{\dimexpr 0.180000\linewidth-5.4000pt\relax}>{\raggedright\arraybackslash}p{\dimexpr 0.210000\linewidth-6.3000pt\relax}>{\raggedright\arraybackslash}p{\dimexpr 0.270000\linewidth-8.1000pt\relax}>{\raggedright\arraybackslash}p{\dimexpr 0.160000\linewidth-4.8000pt\relax}>{\raggedright\arraybackslash}p{\dimexpr 0.090000\linewidth-2.7000pt\relax}>{\raggedright\arraybackslash}p{\dimexpr 0.090000\linewidth-2.7000pt\relax}@{}}\toprule
\textbf{Stage} & \textbf{Initial con\-figu\-ra\-tions} & \textbf{Exploration conditions} & \textbf{Run length} & \textbf{New teacher-labelled con\-figu\-ra\-tions} & \textbf{New DFT reference con\-figu\-ra\-tions} \\\midrule
Initial framework references\newline (iteration 0) & Framework/\allowbreak guest reference structures & DFT references & --- & 0 & 30 \\
Framework/\allowbreak guest temperature exploration\newline (iteration 0--9) & A--F frameworks; 40 atoms + 0/\allowbreak 1/\allowbreak 2 CO\textsubscript{2} & NVT; 100, 300, 500, 800 K & 10 \ensuremath{\rightarrow} 200 ps & 1,016 & 280 \\
Empty-framework feedback\newline (iteration 10--12) & A--F empty frameworks, 40 atoms & NVT; 100, 300, 500, 800 K & 200 ps & 180 & 40 \\
Larger-cell dynamics\newline (iteration 12 r2--r7) & 160-atom frameworks + 0/\allowbreak 4/\allowbreak 8 CO\textsubscript{2} & NVT; 100, 300, 500, 800 K initially; 300, 500 K for longer runs & 10/\allowbreak 30/\allowbreak 50/\allowbreak 100/\allowbreak 150/\allowbreak 200 ps & 540 & 0 \\
Strained-framework relaxation\newline (iteration 13 r0--r1) & A--F strained frameworks + 0/\allowbreak 1/\allowbreak 2 CO\textsubscript{2} & Fixed-cell FIRE; force criterion 0.02 eV Å\textsuperscript{\ensuremath{-}1} & \ensuremath{\leq}2,000 optimiser steps & 346 & 0 \\
Relaxed-framework short dynamics\newline (iteration 13 r2--r4) & Relaxed framework/\allowbreak guest structures & NVT; 100, 300, 500 K & 5 ps & 540 & 0 \\
Hydroxyl-orientation expansion\newline (iteration 13 r5--r7) & Perturbed A--F empty frameworks & Fixed-cell FIRE; force criterion 0.02 eV Å\textsuperscript{\ensuremath{-}1} & \ensuremath{\leq}2,000 optimiser steps & 360 & 0 \\
Isolated-molecule references\newline (iteration 14) & 20 isolated CO\textsubscript{2} geometries & DFT references & --- & 0 & 20 \\
Adsorption-contact relaxation\newline (iteration 14 r0--r2) & 40-atom frameworks + 1/\allowbreak 2 CO\textsubscript{2} & Fixed-cell FIRE; force criterion 0.02 eV Å\textsuperscript{\ensuremath{-}1} & \ensuremath{\leq}2,000 optimiser steps & 360 & 0 \\
Variable-loading exploration\newline (iteration 14 r3--r7) & 160-atom frameworks; variable CO\textsubscript{2} loading & GCMC--MD (NVT MD segments); 300, 500 K; 0.01, 0.05, 0.1, 0.5, 1 bar & 4/\allowbreak 8 ps MD; 40,000/\allowbreak 80,000 MC attempts & 885 & 0 \\
Isolated-atom references\newline (iteration 15) & Al, C, H and O atoms & DFT references & --- & 0 & 4 \\
Larger-framework insertion exploration\newline (iteration 15 r0--r1) & A--F empty frameworks, 160 atoms & Widom \cite{si_widom1963} CO\textsubscript{2} proposals & 10,000 proposals per host & 240 & 0 \\
Small-framework insertion feedback\newline (iteration 15--19) & A--F empty frameworks, 40 atoms & Widom CO\textsubscript{2} proposals & 10,000 proposals per host & 1,080 & 120 \\
Total &  &  &  & 5,547 & 494 \\
\bottomrule\end{tabular}
\par\vspace{2mm}Final student training set: 6,041 configurations.
\end{minipage}\par\medskip

\clearpage

\noindent\begin{minipage}{\linewidth}\setstretch{1}\fontsize{9}{10.8}\selectfont
\textbf{Supplementary Table 4 | Configuration-exploration costs by system and configuration-generation method.}\par\vspace{3mm}
\setlength{\tabcolsep}{3pt}\renewcommand{\arraystretch}{1.12}
\begin{tabular}{@{}>{\raggedright\arraybackslash}p{\dimexpr 0.180000\linewidth-3.2400pt\relax}>{\raggedright\arraybackslash}p{\dimexpr 0.420000\linewidth-7.5600pt\relax}>{\raggedright\arraybackslash}p{\dimexpr 0.200000\linewidth-3.6000pt\relax}>{\raggedright\arraybackslash}p{\dimexpr 0.200000\linewidth-3.6000pt\relax}@{}}\toprule
\textbf{System} & \textbf{Configuration-generation method} & \textbf{Recorded student time (h)} & \textbf{Estimated teacher time (h)} \\\midrule
Al\textsubscript{2}O\textsubscript{3} & NVT & 3.79 & 63.67 \\
Al\textsubscript{2}O\textsubscript{3} & Temperature-programmed NVT & 1.18 & 75.37 \\
Al\textsubscript{2}O\textsubscript{3} & Temperature/\allowbreak cell-programmed NVT & 0.44 & 22.04 \\
Al\textsubscript{2}O\textsubscript{3} & NPT & 2.36 & 143.04 \\
Al\textsubscript{2}O\textsubscript{3} & Fixed-cell FIRE relaxation & 0.03 & 0.02 \\
Al\textsubscript{2}O\textsubscript{3} & System total & 7.79 & 304.14 \\
YSZ & NVT & 1.89 & 64.53 \\
YSZ & NPT & 3.06 & 84.82 \\
YSZ & Temperature-programmed NPT & 1.84 & 52.77 \\
YSZ & System total & 6.79 & 202.13 \\
MIL-120(Al) & NVT & 6.41 & 141.96 \\
MIL-120(Al) & Fixed-cell FIRE relaxation & 0.38 & 0.22 \\
MIL-120(Al) & Widom insertion & 0.85 & 1.56 \\
MIL-120(Al) & GCMC--MD & 4.19 & 86.09 \\
MIL-120(Al) & System total & 11.83 & 229.82 \\
\bottomrule\end{tabular}
\par\vspace{2mm}Times are summed over atom-count and output-cadence modes within each configuration-generation method. Student entries retain historical execution settings; teacher entries use the selected parallel configurations. System totals are calculated before rounding.
\end{minipage}\par\medskip

\clearpage

\section*{Supplementary Note 2 | Controlled teacher-feedback experiments}

\noindent\begin{minipage}{\linewidth}\centering
\includegraphics[width=180.00mm,height=146.56mm,keepaspectratio]{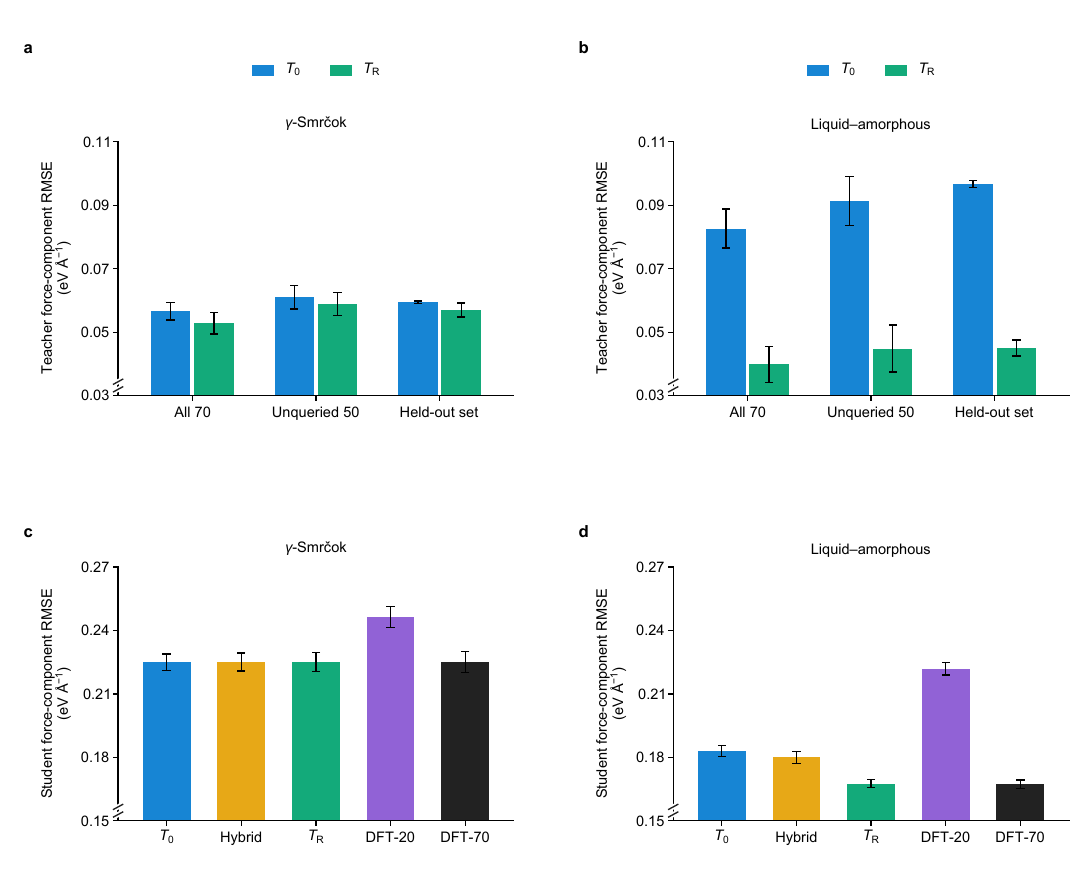}
\par\vspace{2mm}{\fontsize{10}{12}\selectfont\raggedright\textbf{Supplementary Fig. 1 | Teacher and student force errors in the controlled experiments.} a,b, Initial ($T_0$) and updated ($T_R$) teacher force-component RMSE on seventy student-training geometries, fifty additions excluded from teacher adaptation, and held-out configurations. c,d, Held-out student RMSE. Columns show $\gamma$-Smr\v{c}ok (446 held-out configurations) and liquid--amorphous Al\textsubscript{2}O\textsubscript{3} (363). Bars and error bars show means and sample SDs across five configuration-selection repeats.\par}
\end{minipage}\par

\clearpage

\section*{Supplementary Note 3 | QBC and support-guided routing}

\noindent\begin{minipage}{\linewidth}\centering
\includegraphics[width=180.00mm,height=146.56mm,keepaspectratio]{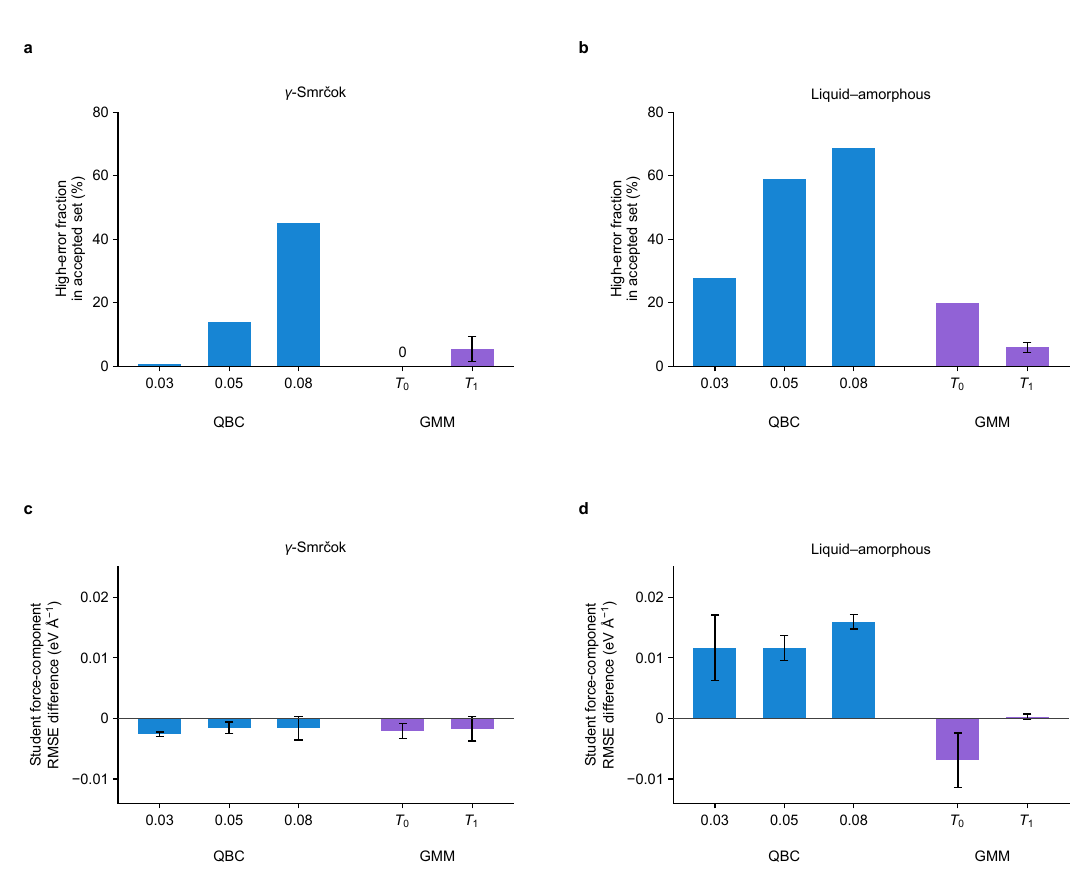}
\par\vspace{2mm}{\fontsize{10}{12}\selectfont\raggedright\textbf{Supplementary Fig. 2 | Configuration selection for teacher labelling and student accuracy.} a,b, High-error fractions among configurations accepted for teacher labelling by QBC ($T_0$; $\sigma$ thresholds in eV Å\textsuperscript{\ensuremath{-}1}) and GMM ($T_0$/$T_1$; $p>0.1$). High error denotes configuration force-component RMSE > 0.05 eV Å\textsuperscript{\ensuremath{-}1}. c,d, Held-out student RMSE differences (students trained on teacher labels minus students trained entirely on DFT labels; identical training geometries). Columns show $\gamma$-Smr\v{c}ok and liquid--amorphous Al\textsubscript{2}O\textsubscript{3}. Bars and error bars show means and sample SDs across five repeats; initial routing is evaluated once. $T_1$ routing excludes the ten newly queried DFT configurations.\par}
\end{minipage}\par

\clearpage

\section*{Supplementary Note 4 | DICE distillation versus direct fine-tuning}

\noindent\begin{minipage}{\linewidth}\setstretch{1}\fontsize{9}{10.8}\selectfont
\textbf{Supplementary Table 5 | Dynamical stability of the final student potentials.}\par\vspace{3mm}
\setlength{\tabcolsep}{3pt}\renewcommand{\arraystretch}{1.12}
\begin{tabular}{@{}>{\raggedright\arraybackslash}p{\dimexpr 0.150000\linewidth-2.7000pt\relax}>{\raggedright\arraybackslash}p{\dimexpr 0.350000\linewidth-6.3000pt\relax}>{\raggedright\arraybackslash}p{\dimexpr 0.250000\linewidth-4.5000pt\relax}>{\raggedright\arraybackslash}p{\dimexpr 0.250000\linewidth-4.5000pt\relax}@{}}\toprule
\textbf{System} & \textbf{Simulation conditions} & \textbf{Directly fine-tuned student} & \textbf{DICE-distilled student} \\\midrule
Al\textsubscript{2}O\textsubscript{3} & Liquid; 1,920 atoms; 2400 K; NPT & Dynamical instability at 0.22 ps & Stable through 1 ns \\
10YSZ & 640 atoms; 1800 K; NPT & Dynamical instability at 0.3 ps & Stable through 1 ns \\
MIL-120(Al) & Empty framework; 160 atoms; 500 K; NVT & Dynamical instability at 593 ps & Stable through 1 ns \\
\bottomrule\end{tabular}
\par\vspace{2mm}The prescribed duration is 1 ns. Timesteps are 1 fs for Al\textsubscript{2}O\textsubscript{3} and YSZ and 0.5 fs for MIL-120(Al). The earliest recorded instability time is determined from the last stored frame.
\end{minipage}\par\medskip

\clearpage

\noindent\begin{minipage}{\linewidth}\centering
\includegraphics[width=180.00mm,height=157.43mm,keepaspectratio]{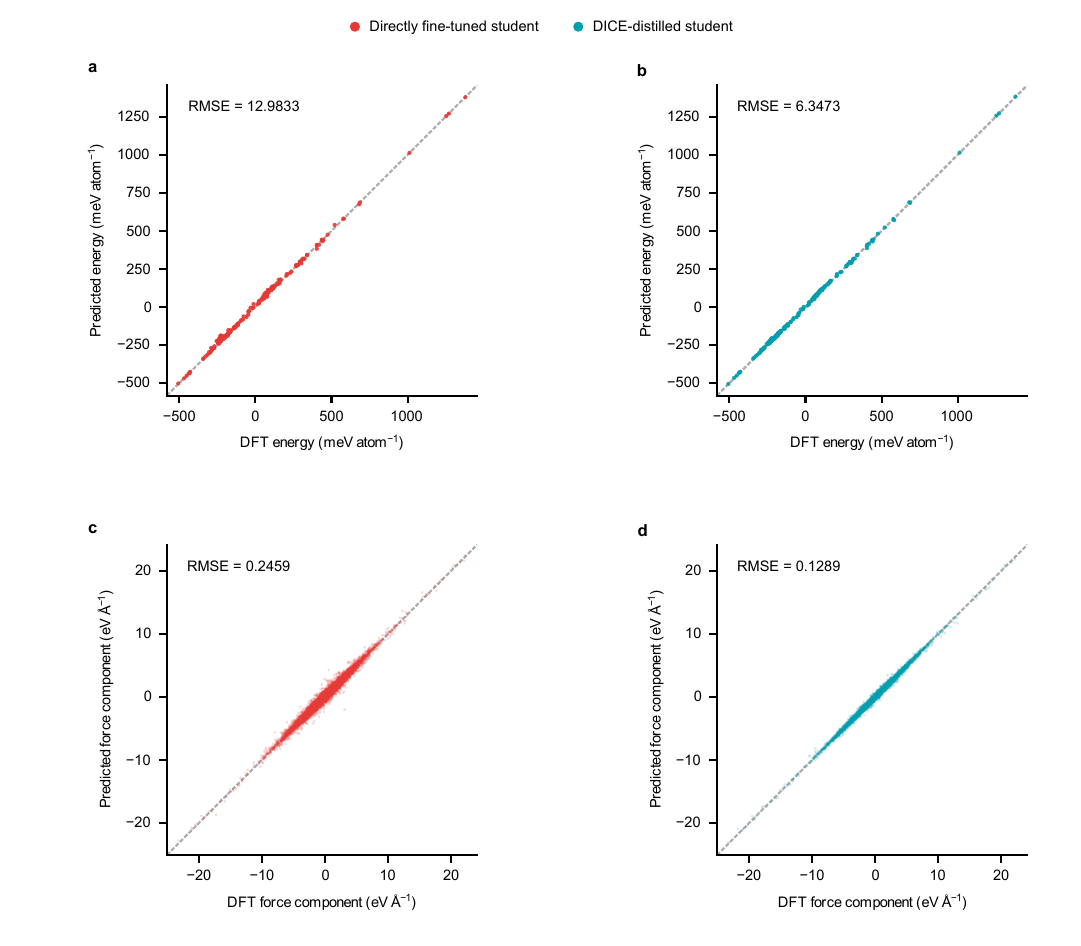}
\par\vspace{2mm}{\fontsize{10}{12}\selectfont\raggedright\textbf{Supplementary Fig. 3 | Al\textsubscript{2}O\textsubscript{3} energy and force parity.} a,b, Per-atom energies; c,d, Cartesian force components for directly fine-tuned (a,c) and DICE-distilled (b,d) students on 200 Al\textsubscript{2}O\textsubscript{3} test configurations. Energies subtract the common DFT mean; dashed lines indicate equality.\par}
\end{minipage}\par

\clearpage

\noindent\begin{minipage}{\linewidth}\centering
\includegraphics[width=180.00mm,height=157.43mm,keepaspectratio]{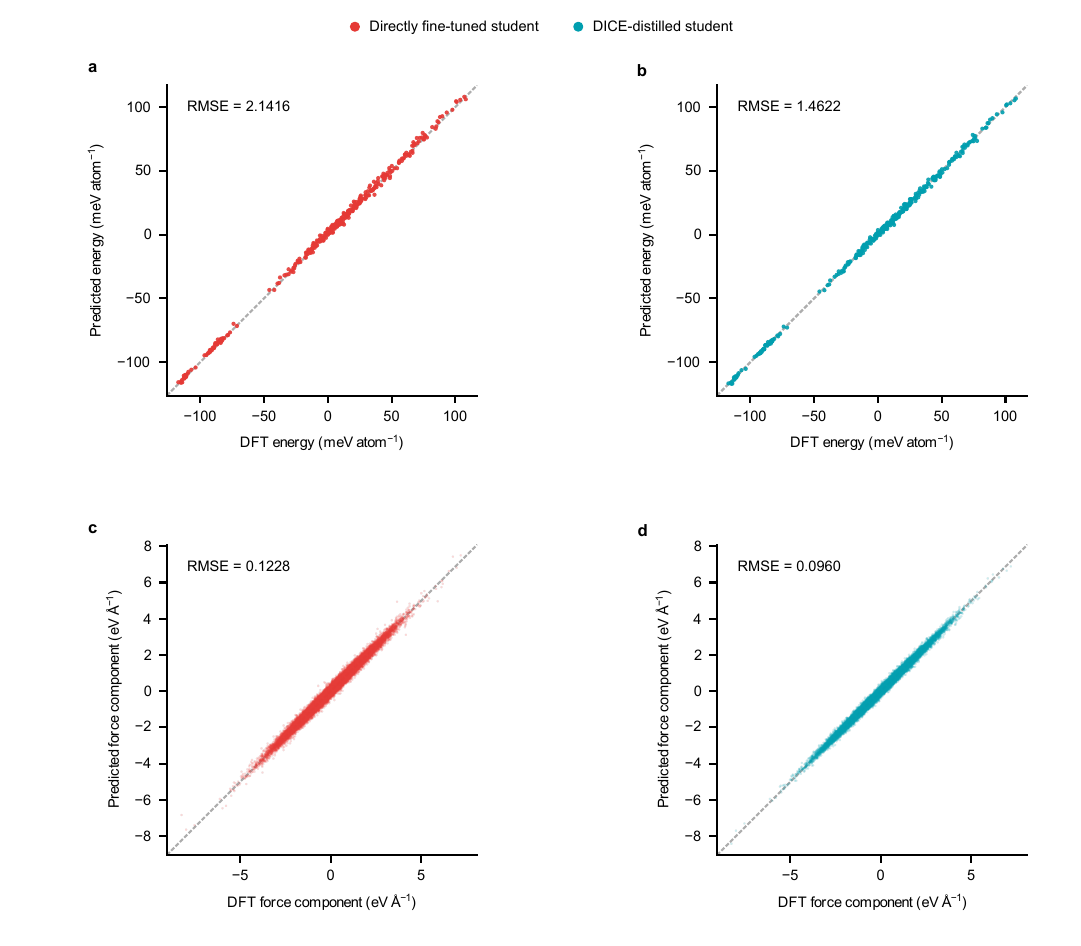}
\par\vspace{2mm}{\fontsize{10}{12}\selectfont\raggedright\textbf{Supplementary Fig. 4 | YSZ energy and force parity.} a,b, Per-atom energies; c,d, Cartesian force components for directly fine-tuned (a,c) and DICE-distilled (b,d) students on 330 YSZ test configurations. Energies subtract the common composition-specific DFT mean.\par}
\end{minipage}\par

\clearpage

\noindent\begin{minipage}{\linewidth}\centering
\includegraphics[width=180.00mm,height=157.43mm,keepaspectratio]{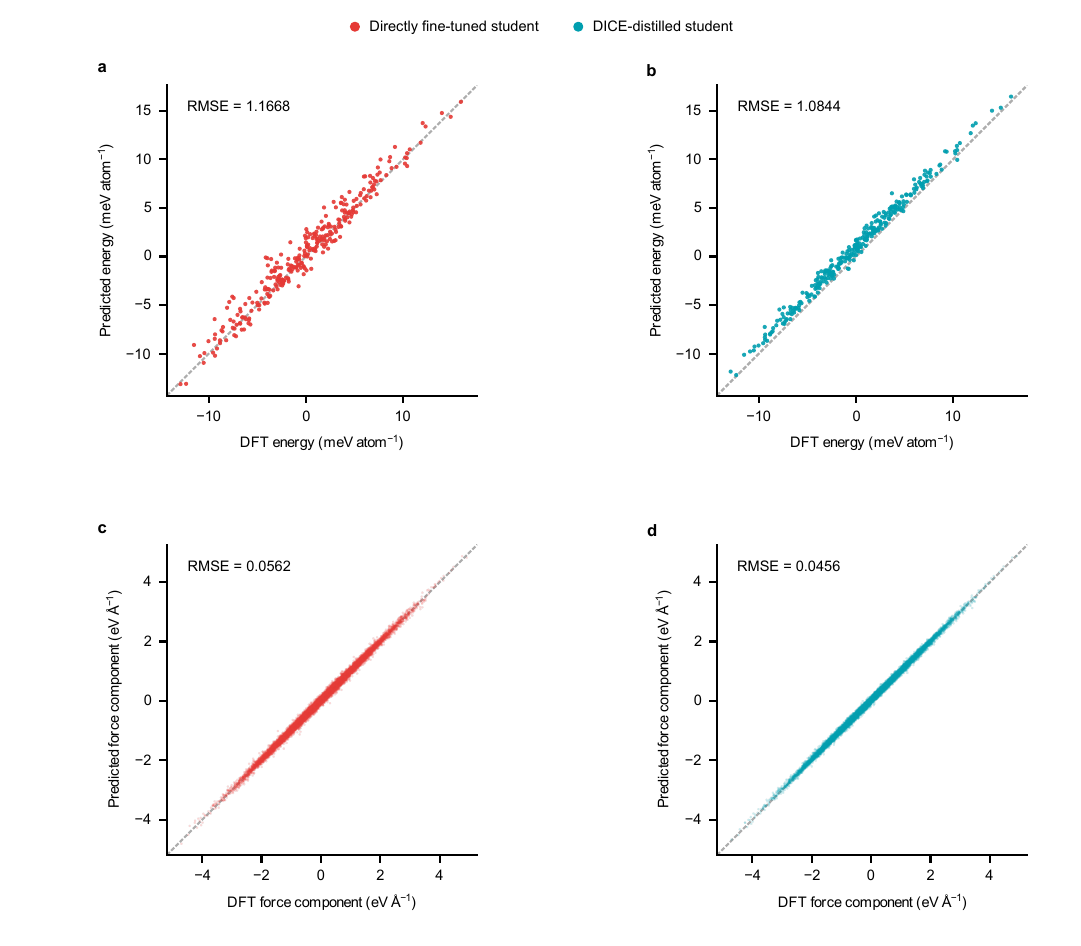}
\par\vspace{2mm}{\fontsize{10}{12}\selectfont\raggedright\textbf{Supplementary Fig. 5 | MIL-120(Al) energy and force parity.} a,b, Per-atom energies; c,d, Cartesian force components for directly fine-tuned (a,c) and DICE-distilled (b,d) students on 300 MIL-120(Al) test configurations with 40-atom frameworks. Energies subtract the common DFT mean.\par}
\end{minipage}\par

\clearpage

\noindent\begin{minipage}{\linewidth}\centering
\includegraphics[width=180.00mm,height=167.30mm,keepaspectratio]{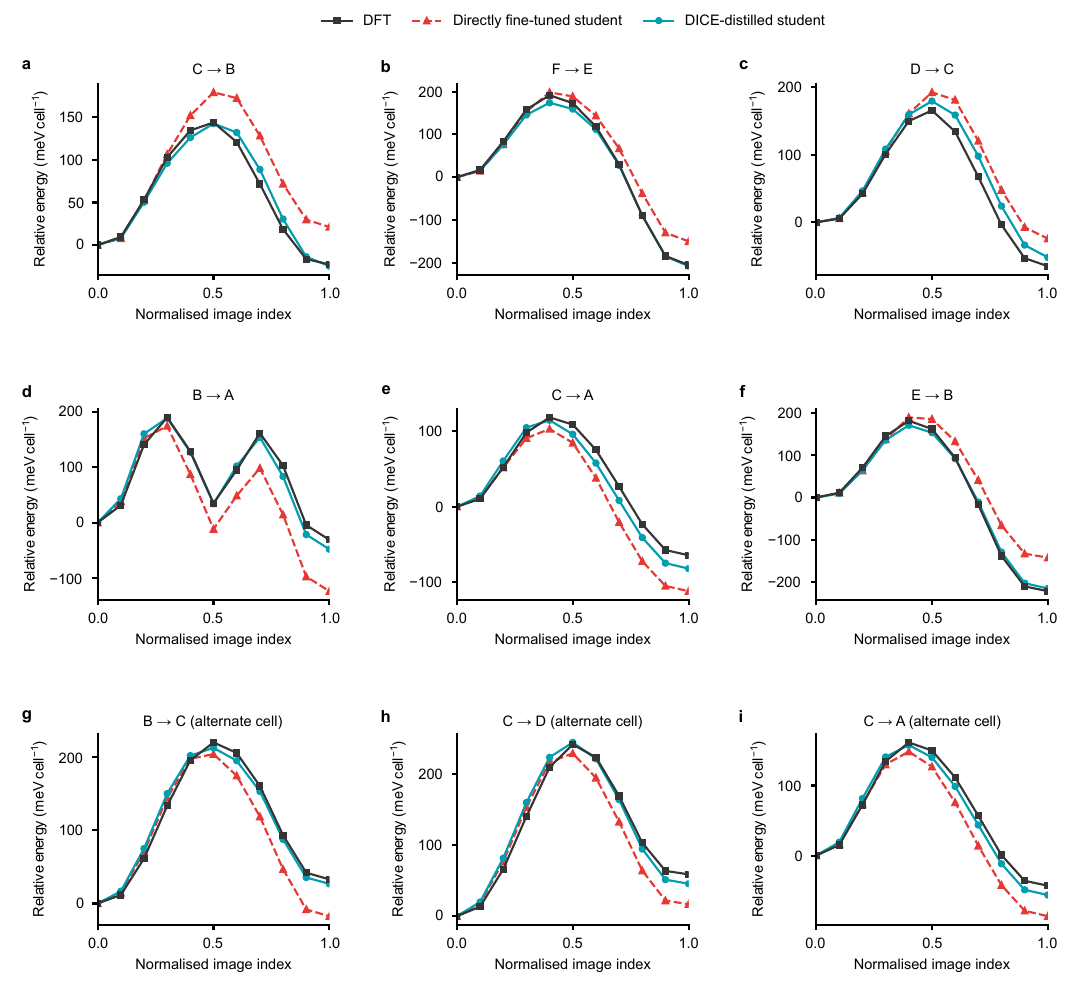}
\par\vspace{2mm}{\fontsize{10}{12}\selectfont\raggedright\textbf{Supplementary Fig. 6 | MIL-120(Al) hydroxyl-rearrangement paths on common DFT geometries.} DFT and both students evaluated on identical geometries for nine empty-framework paths \cite{si_dice_ref_17}. Each method's initial-image energy is zero; ``alternate cell'' denotes an alternative fixed cell.\par}
\end{minipage}\par

\clearpage

\noindent\begin{minipage}{\linewidth}\centering
\includegraphics[width=180.00mm,height=115.00mm,keepaspectratio]{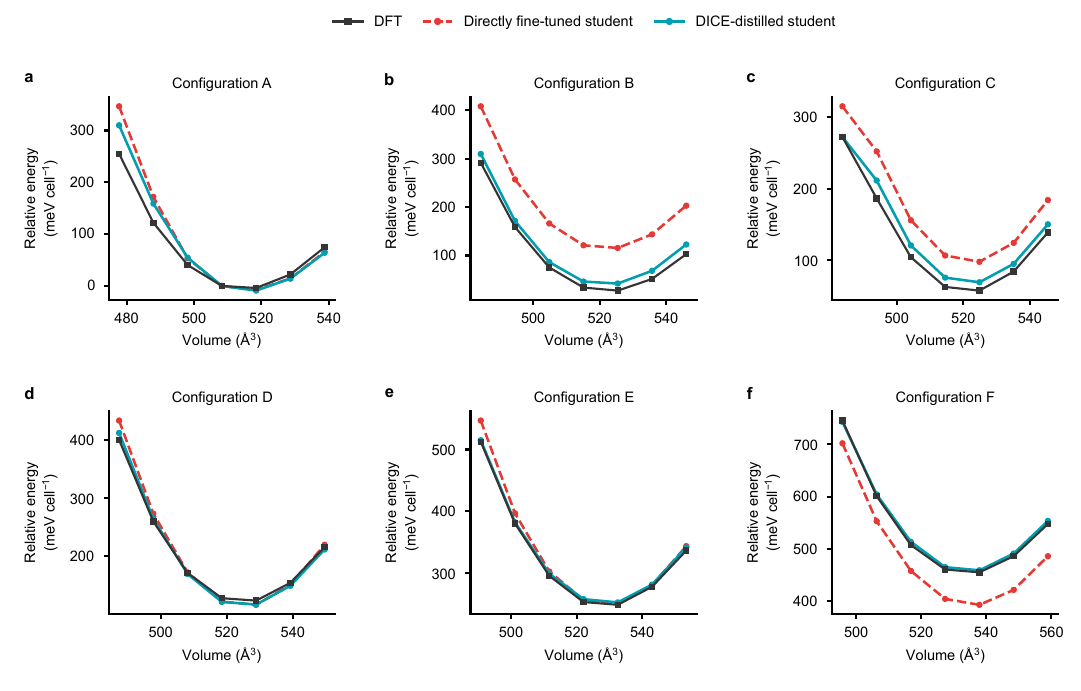}
\par\vspace{2mm}{\fontsize{10}{12}\selectfont\raggedright\textbf{Supplementary Fig. 7 | Equations of state for six MIL-120(Al) framework configurations.} a--f, Configurations A--F \cite{si_dice_ref_17} evaluated on common DFT geometries; each method uses its A-configuration energy at volume ratio 1.0 as zero.\par}
\end{minipage}\par

\clearpage

\section*{Supplementary Note 5 | Validation of physical properties}

\subsection*{S5.1 Al\textsubscript{2}O\textsubscript{3} relaxation and phonon calculations}

\noindent For the supplementary equilibrium comparison, eleven Al\textsubscript{2}O\textsubscript{3} prototypes are fully relaxed independently with PBEsol DFT and the DICE-distilled student; each method uses its relaxed $\alpha$-phase energy as zero.\par

Reference cells and displacement geometries for $\alpha$-, $\kappa$- and $\theta$-Al\textsubscript{2}O\textsubscript{3} use NIMS PhononDB records mp-1143, mp-2254 and mp-7048, respectively. All models use identical displacement geometries and force-constant processing, without non-analytical corrections.\par

\subsection*{S5.2 Solid--liquid coexistence analysis}

\noindent Uncertainty in the mean interface velocity is estimated by hierarchical bootstrap resampling of trajectory replicates and contiguous time blocks within each trajectory.\par

\subsection*{S5.3 YSZ diffusion and thermal-expansion analysis}

\noindent Lattice parameters and thermal expansion of nominal 8YSZ and 10YSZ are calculated by NPT heating at 0.1 K ps\textsuperscript{\ensuremath{-}1} using fluorite cells containing 864 cations, with exact yttria contents of 8.000 and 10.064 mol\%, respectively. The equivalent cubic lattice parameter is $a=(V/216)^{1/3}$. Linear expansion coefficients are obtained by equal-weight fits of 10 K-bin mean $\ln(V^{1/3})$ against measured temperature.\par

Normalised diffusivity ratios and their pointwise 95\% confidence intervals are obtained from 10,000 bootstrap resamples of complete structural preparations, recalculating each condition mean and the 5.8YSZ denominator at the same temperature.\par

\subsection*{S5.4 MIL-120(Al) pathway optimisation}

\noindent Each path contains eleven images. Path optimisation uses the improved-tangent formulation and fast inertial relaxation engine (FIRE).\par

\clearpage

\noindent\begin{minipage}{\linewidth}\centering
\includegraphics[width=180.00mm,height=105.01mm,keepaspectratio]{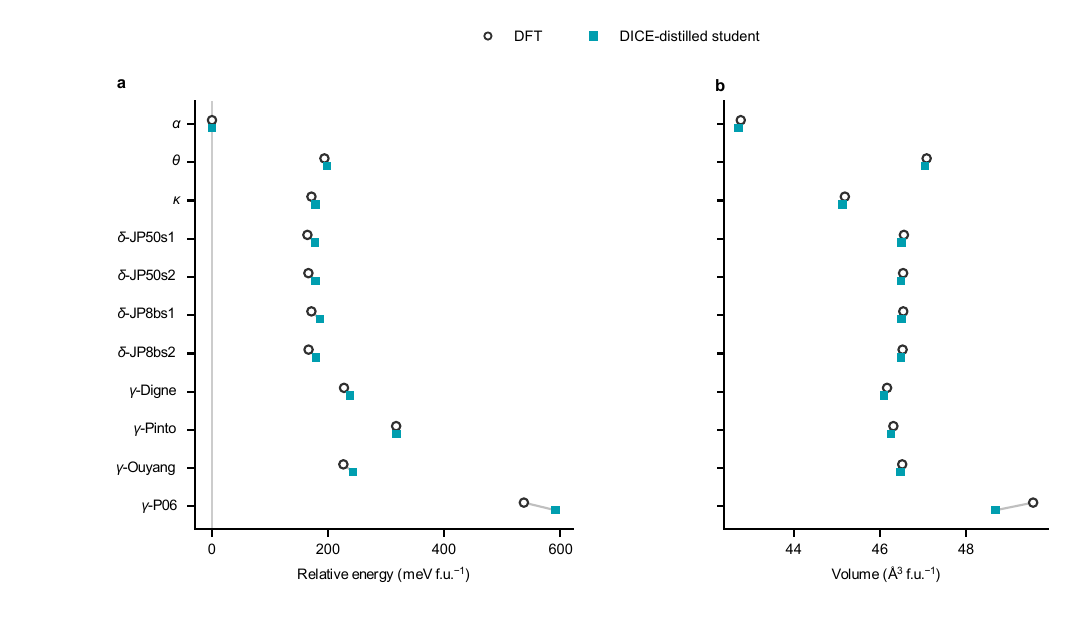}
\par\vspace{2mm}{\fontsize{10}{12}\selectfont\raggedright\textbf{Supplementary Fig. 8 | Relaxed energies and volumes of eleven Al\textsubscript{2}O\textsubscript{3} prototypes.} a,b, Relative energies and equilibrium volumes from independent full relaxations with PBEsol DFT and the DICE-distilled student, expressed per formula unit. Each method's relaxed $\alpha$-phase energy is zero.\par}
\end{minipage}\par

\clearpage

\noindent\begin{minipage}{\linewidth}\centering
\includegraphics[width=180.00mm,height=167.01mm,keepaspectratio]{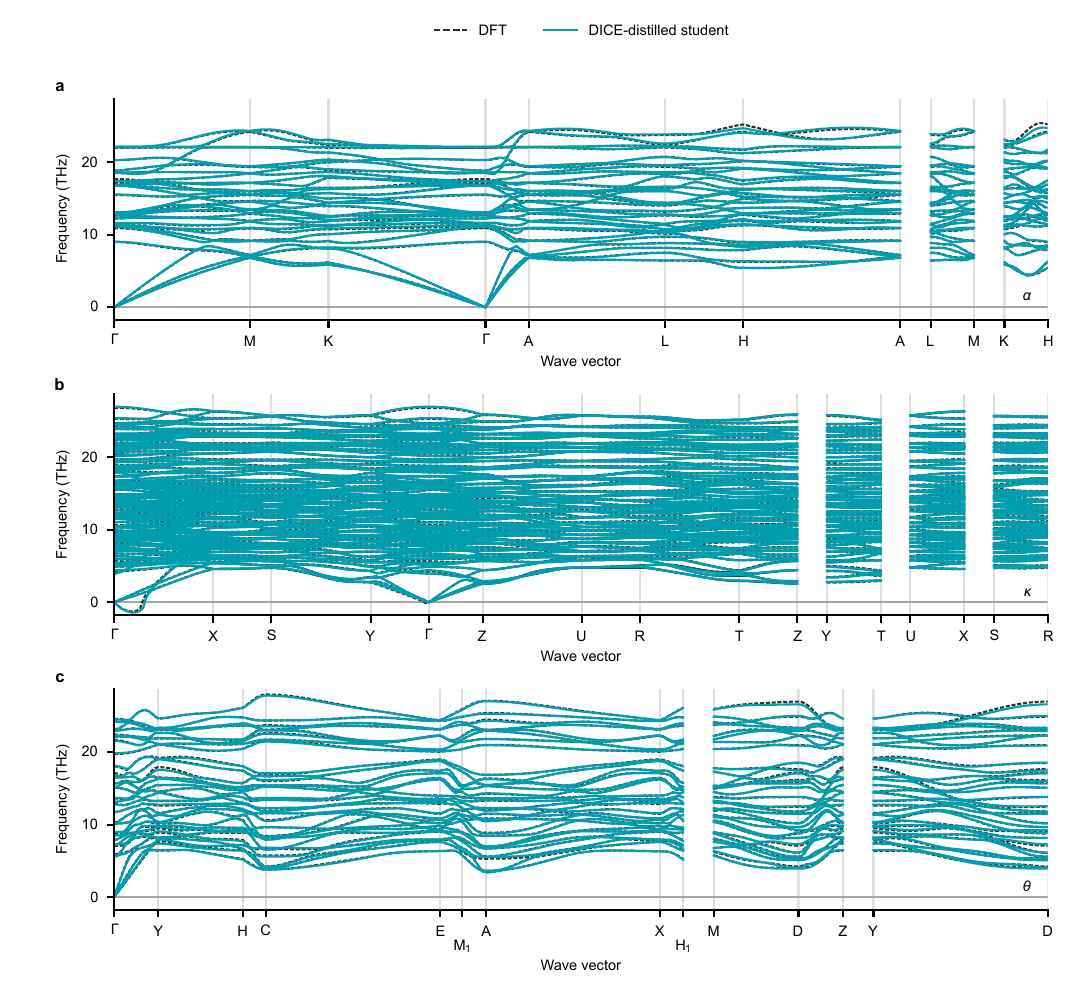}
\par\vspace{2mm}{\fontsize{10}{12}\selectfont\raggedright\textbf{Supplementary Fig. 9 | Phonon dispersions of $\alpha$-, $\kappa$- and $\theta$-Al\textsubscript{2}O\textsubscript{3}.} a--c, DFT and DICE phonon dispersions for $\alpha$-, $\kappa$- and $\theta$-Al\textsubscript{2}O\textsubscript{3}, respectively, using identical reference cells and displacement geometries.\par}
\end{minipage}\par

\clearpage

\noindent\begin{minipage}{\linewidth}\centering
\includegraphics[width=180.00mm,height=81.98mm,keepaspectratio]{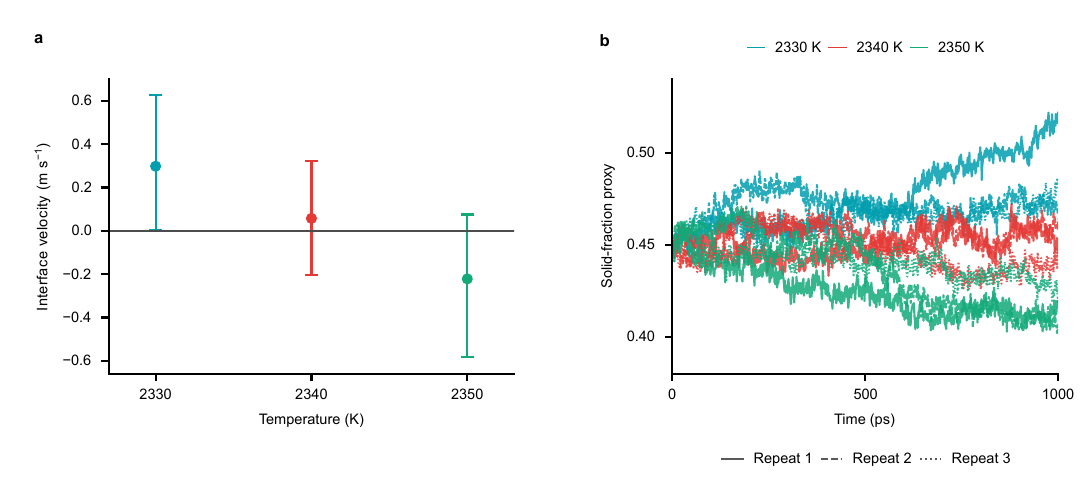}
\par\vspace{2mm}{\fontsize{10}{12}\selectfont\raggedright\textbf{Supplementary Fig. 10 | Interface velocities and finite-time coexistence histories.} a, Mean interface velocities and 95\% hierarchical trajectory/block bootstrap intervals from three 1 ns trajectories per temperature; positive velocity indicates solid growth. b, Solid-fraction-proxy histories at 2330, 2340 and 2350 K; colours indicate temperature and line styles identify trajectories.\par}
\end{minipage}\par

\clearpage

\noindent\begin{minipage}{\linewidth}\centering
\includegraphics[width=180.00mm,height=91.98mm,keepaspectratio]{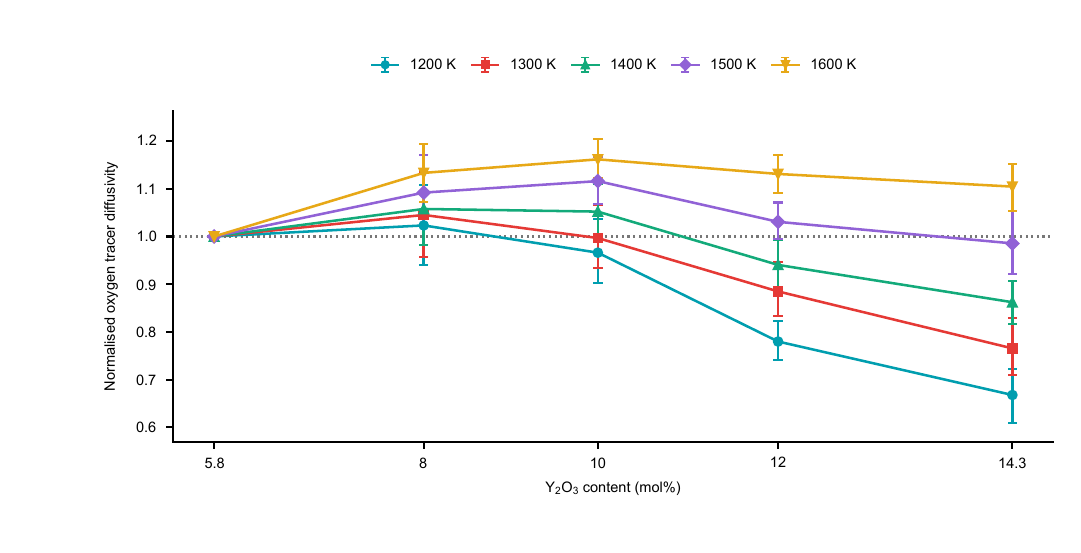}
\par\vspace{2mm}{\fontsize{10}{12}\selectfont\raggedright\textbf{Supplementary Fig. 11 | Composition dependence of normalised oxygen tracer diffusivity.} Oxygen tracer diffusivities at 1200--1600 K, normalised by the 5.8YSZ mean at each temperature. Points show ratios of means from four independent structural preparations; error bars show pointwise 95\% bootstrap confidence intervals with complete preparations as resampling units.\par}
\end{minipage}\par

\clearpage

\noindent\begin{minipage}{\linewidth}\centering
\includegraphics[width=180.00mm,height=165.01mm,keepaspectratio]{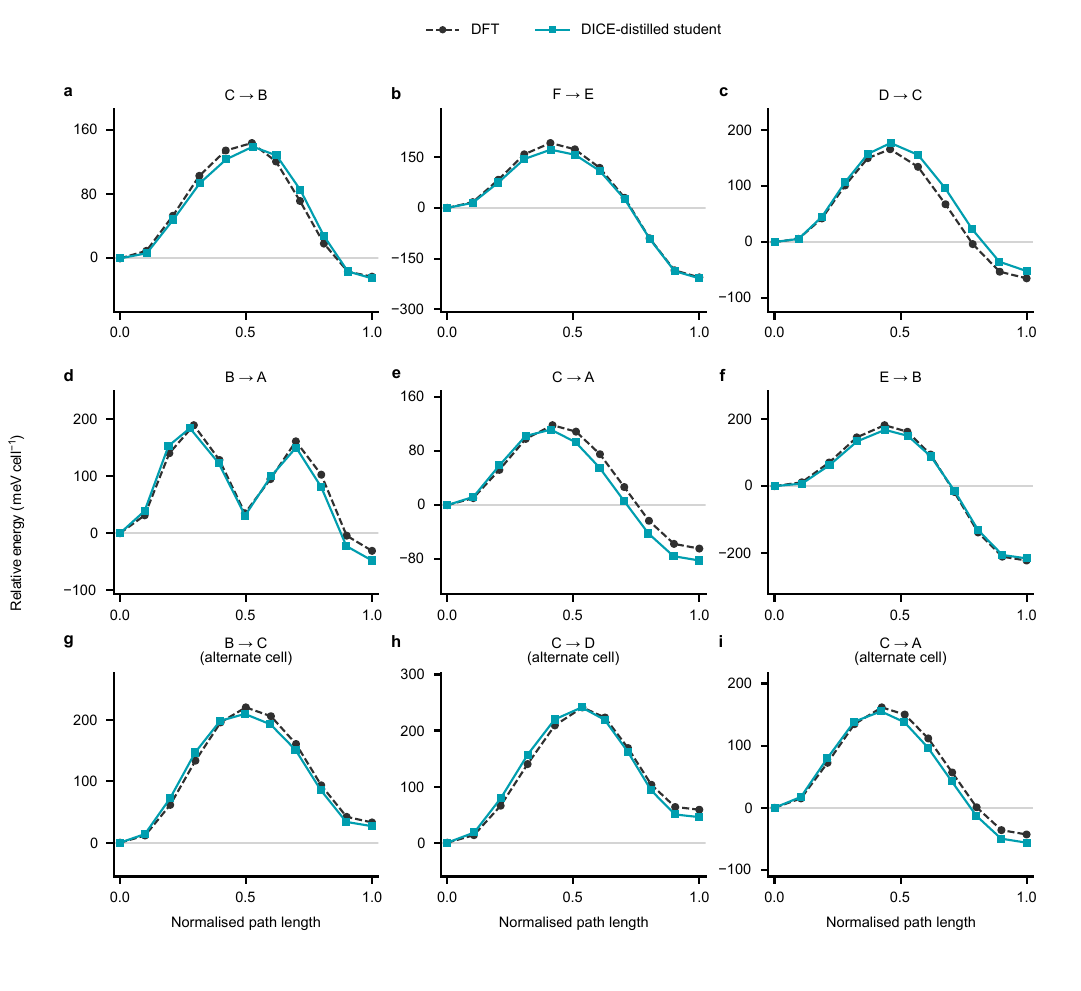}
\par\vspace{2mm}{\fontsize{10}{12}\selectfont\raggedright\textbf{Supplementary Fig. 12 | Model-optimised hydroxyl-rearrangement pathways in MIL-120(Al).} Nine CI-NEB paths optimised independently with DFT and the DICE-distilled student, with fixed DFT endpoints and cells. Each path uses its initial-image energy as zero and cumulative minimum-image displacement normalised by its total length.\par}
\end{minipage}\par

\clearpage

\noindent\begin{minipage}{\linewidth}\centering
\includegraphics[width=180.00mm,height=87.98mm,keepaspectratio]{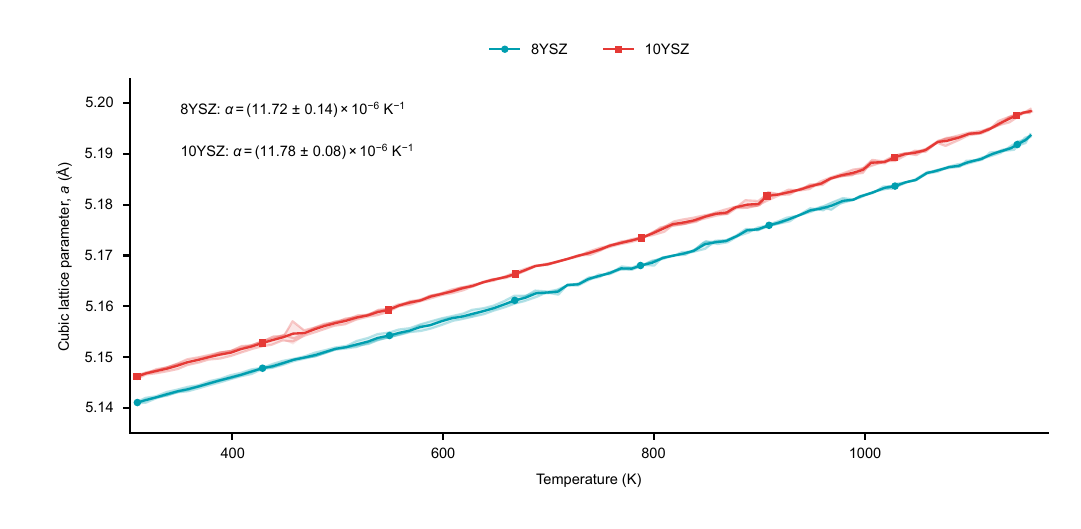}
\par\vspace{2mm}{\fontsize{10}{12}\selectfont\raggedright\textbf{Supplementary Fig. 13 | Lattice parameters and thermal expansion of YSZ.} Equivalent cubic lattice parameters during NPT heating. Thin curves show three independent preparations; thick curves and shading show means and sample SDs. Annotated expansion coefficients are fitted over 303.15--1173.15 K and summarised as mean \ensuremath{\pm} sample SD across preparations. Experimental TMA coefficients over the same interval are 10.31 \ensuremath{\times} 10\textsuperscript{\ensuremath{-}6} and 10.38 \ensuremath{\times} 10\textsuperscript{\ensuremath{-}6} K\textsuperscript{\ensuremath{-}1} for 8YSZ and 10YSZ, respectively \cite{si_si_gao_2017_thermal,si_si_gao_2016_thesis}.\par}
\end{minipage}\par

\clearpage

\clearpage

\renewcommand{\refname}{Supplementary References}

\begingroup\setstretch{1.05}\fontsize{11}{13}\selectfont\endgroup